\documentclass[12pt]{article}
\usepackage[dvipsnames,table]{xcolor}
\usepackage{booktabs,siunitx}
\usepackage{color, verbatim}
\usepackage{latexsym}
\usepackage{amssymb}
\usepackage{amsmath}
\usepackage{graphicx}
\usepackage{arydshln}
\usepackage{wasysym}
\usepackage{adjustbox}
\usepackage{easybmat}
\usepackage{bbm}
\usepackage{cite}
\usepackage{slashed}
\usepackage{bm}
\usepackage{multirow} 
\usepackage{rotating}
\usepackage{mathtools}
\usepackage[inline]{enumitem}
\usepackage{lscape}
\usepackage[normalem]{ulem}
\usepackage{booktabs}
\usepackage{colortbl}
\usepackage{pstricks}
\usepackage{caption, subcaption}
\usepackage{lineno}
\usepackage{indentfirst}

\newcommand{\gtick}{\textcolor{green}{\mathbf{\checkmark}}}
\newcommand{\otick}{\textcolor{orange}{\mathbf{\checkmark}}}
\newcommand{\rtick}{\textcolor{red}{\text{\sffamily X}}}

\newcommand{\blname}{\cellcolor{blue!20}}
\newcommand{\blop}{\cellcolor{blue!10}}
\newcommand{\bhname}{\cellcolor{blue!25}}
\newcommand{\bhop}{\cellcolor{blue!15}}
\newcommand{\bclass}{\cellcolor{blue!40}}
\newcommand{\rlname}{\cellcolor{orange!20}}
\newcommand{\rlop}{\cellcolor{orange!10}}
\newcommand{\rhname}{\cellcolor{orange!25}}
\newcommand{\rhop}{\cellcolor{orange!15}}
\newcommand{\rclass}{\cellcolor{orange!40}}
\newcommand{\glname}{\cellcolor{green!20}}
\newcommand{\glop}{\cellcolor{green!10}}
\newcommand{\ghname}{\cellcolor{green!25}}
\newcommand{\ghop}{\cellcolor{green!15}}
\newcommand{\gclass}{\cellcolor{green!40}}
\newcommand{\olname}{\cellcolor{red!20}}
\newcommand{\olop}{\cellcolor{red!10}}

\newcommand{\oclass}{\cellcolor{red!40}}
\usepackage{hyperref}
\hypersetup{
	colorlinks = true,
	linkcolor = blue,
	anchorcolor = blue,
	citecolor = blue,
	filecolor = blue,
	urlcolor = magenta,
	linktocpage=true
}
\usepackage{url}

\newcommand{\ii}{\ensuremath{\mathrm{i}}}
\newcommand{\order}[1]{$1/\Lambda^{#1}$}

\usepackage[symbol]{footmisc}

\renewcommand{\theequation}{\thesection.\arabic{equation}}
\makeatletter
\@addtoreset{equation}{section}
\g@addto@macro\bfseries{\boldmath}
\newcommand\Label[1]{&\refstepcounter{equation}(\theequation)\ltx@label{#1}&}
\makeatother

\allowdisplaybreaks

\begin{document}
	
	\thispagestyle{empty}
	\begin{flushright}
	\end{flushright}
	\vspace{0.8cm}
	
	\begin{center}
		{\Large\sc Renormalisation of SMEFT bosonic interactions \\[8pt] up to dimension eight by LNV operators
		}
		\vspace{0.8cm}
		
		\textbf{ Supratim Das Bakshi$^{\,a}\footnote[1]{\href{sdb.ugr.es}{sdb@ugr.es}}$ and \'Alvaro D\'iaz-Carmona$^{\,a}\footnote[2]{\href{aldiaz@ugr.es}{aldiaz@ugr.es}}$ }\\
		\vspace{.5cm}
		{\em {$^a$ CAFPE and Departamento de F\'isica Te\'orica y del Cosmos,
				Universidad de Granada, Campus de Fuentenueva, E--18071 Granada, Spain}}\\[1cm]
	\end{center}
	\begin{abstract}
		We present the renormalisation group running of the bosonic operators of the Standard Model effective field theory (SMEFT) by the Lepton Number Violating operators (LNVs) at 1-loop order up to $\mathcal{O}(v^4/\Lambda^4)$, with $v \sim  246$ GeV as the electroweak scale and $\Lambda$ as the SMEFT cut-off. Using these relations with the positivity bounds on Wilson coefficients of $\phi^4 D^4$ class, we derive sign constraints on the Wilson coefficients of LNV operators, for models where $\phi^4 D^4$  operators do not appear at tree-level. We inspect these constraints for the LNV Wilson coefficients generated from matching the Type-I and III seesaw models to SMEFT up to dimension seven at tree-level. We also exhibit the unique bounds induced by the T-parameter on LNVs.
	\end{abstract} 
	
	\newpage
	
	\tableofcontents

	\section{Introduction}
	The Standard Model Effective Field Theory is a robust framework to parametrise New Physics (NP), as it extends the Standard Model (SM) with effective interactions accounting for the experimental observations supporting physics beyond SM (BSM) \cite{Super-Kamiokande:1998kpq}. 
	These effective interactions formed out of the SM particle content respect the gauge symmetry $ SU(3)_C\otimes SU(2)_L\otimes U(1)_Y $ and are suppressed by the SMEFT cut-off scale. In the presence of these effective operators, the modifications to the observables are parameterised in terms of the SMEFT Wilson coefficients (WCs), which encapsulate the deviations from the SM \cite{PhysRevLett.43.1566,Brivio:2017vri,Ellis:2018gqa,Ellis:2020ljj,Anisha:2020ggj,Corbett:2021eux,Chala:2021juk,Anisha:2021hgc}.
	
	To analyse NP interactions via the observables defined at separated energy scales, one needs to consistently relate WC from these scales. In the present scenario, where the NP is apparently beyond the reach of current searches, it is sensible to assume that the NP scale is much larger than that of the Electroweak (EW) Theory. In a generic (top-down) EFT scenario where NP models are matched to SMEFT, the WCs of effective operators defined at high scale encapsulate the NP interactions and scale. The WCs encode the couplings featured by a theory defined at a certain energy level. Renormalisation Group Equations (RGEs) are key to expressing these WCs at energies accessible to the measurements. We construct these relations for SMEFT, and for this, we consider a finite experimental resolution, which facilitates a truncation on the infinite tower of effective operators suppressed by according powers of cut-off ($\Lambda$).
	
	Recent studies indicate that the SMEFT truncation at order \order{2} is insufficient for making consistent predictions in several cases.
	For instance, dimension-8 interactions are leading contributors in many observables \cite{Degrande:2013kka,Azatov:2016sqh,Ellis:2020ljj,Guo:2020lim,Alioli:2020kez}. In the weakly-coupled Ultraviolet (UV) completions of the SMEFT, certain dimension-6 classes of operators are not generated at tree-level \cite{Murphy:2020rsh,Craig:2019wmo,DasBakshi:2020pbf,Bakshi:2021ofj}. The contribution from dimension-8 (and dimension-6 squared) effects is significant when the SMEFT cut-off is chosen relatively low \cite{Contino:2016jqw,Baglio:2020oqu}. Order \order{4} is also needed in certain cases where the observables are measured so precisely that they become sensitive to higher mass dimension operators' effects \cite{Panico:2018hal,Ardu:2021koz,Corbett:2021eux}.
	
	These studies motivate to extend the SMEFT analysis to order \order{4} by including dimension-8 operators effects, which is the method followed in many recent theoretical studies \cite{Chala:2018ari,Hays:2018zze,Alioli:2020kez,Hays:2020scx,Gu:2020ldn,Corbett:2021eux}. Renormalisation up to order \order{4} in the SMEFT from two dimension-6 operators \cite{Chala:2021pll,Helset:2022pde}, and one dimension-8 operator \cite{AccettulliHuber:2021uoa,DasBakshi:2022mwk} have been studied in detail. This paper aims to compute the renormalisation of SMEFT bosonic interactions up to dimension-8 by including the running induced by lepton number violating operators up to dimension-7.  The effects of these dimension-5 and dimension-7 effective operators are suppressed by lepton number violating scale. This scale is  typically assumed to be very high ($\geqslant \mathcal{O}(10^{9})$ GeV) in certain models (for instance, in Type-I seesaw: $M_\nu \simeq M_D (M_N)^{-1} M_D$, taking $ M_D \sim \mathcal{O}(1) $, and $M_\nu$ and $M_N$ are light and heavy neutrino masses respectively). However, there are also models of neutrino mass generation which explore the possibility of lighter LNV scale, for instance, radiative seesaw models \cite{RFigueiredo:2014tap,Law:2013dya}, neutrino texture models \cite{PhysRevD.84.033002,Mitra:2011qr}, etc.\footnote{See also Ref.~\cite{Dong:2011rh}, where the possibility of observing baryon number violating processes in context of top-quark production/decay at hadron colliders is discussed.} Here, we take an agnostic approach on the LNV scale relying upon the robustness of bottom-up approach of EFTs.
	
	With LNV operators included in the RGEs, all the possible contributions to bosonic operators of the SMEFT are known explicitly up to order \order{4} at 1-loop. There are several reasons to approach this problem. For instance, the running of some operators can provide the leading SMEFT corrections to SM predictions of observables in which loop-induced interactions appear at leading order. Also, under certain conditions, constraints on dimension-8 operators can reveal unique bounds on the space of LNV WCs. We discuss one such case where the T-parameter bound is translated to LNV operators.
	
	Dimension-8 Wilson coefficients are constrained by positivity bounds which are restrictions on S-matrix elements deduced from the fundamental axioms -- analyticity, unitarity and crossing symmetry on the underlying theory. \cite{Adams:2006sv,Zhang:2018shp,Bi:2019phv,Remmen:2019cyz,Remmen:2020uze,Bonnefoy:2020yee,Bellazzini:2020cot}. Recently, these bounds for the dimension-8 operators and their validity under RG running have been analysed in detail \cite{Chala:2021wpj,DasBakshi:2022mwk,Li:2022aby}. Following the top-down approach, SMEFT WCs are related to model dependent NP couplings. The RGE effects and the positivity bounds on these WCs impose remarkable restrictions on the NP parameter space on a case-to-case basis. In this work, we show unique constraints on the UV space deduced solely from the RGE computation in the IR (SMEFT in this case) and positivity bounds. We also examine these constraints on two BSMs -- Type-I and III seesaw models -- by matching them to SMEFT up to dimension-7 at tree-level, then validating the consistency of the derived constraints. Related to this, the T-parameter impose restrictions on dimension-8 operators of the $ \phi^6D^2 $ class. Based on the RGE relations computed here, we deduce new restrictions on the LNVs driven by the T-parameter constraint \cite{deblas:2016nqo}. We comment on the complementarity of this outcome with the one deduced from the neutrino mass. We point out that this RGE induced constraint on LNV Wilson coefficients restrict them from arbitrary large (absolute) values which is a flat direction in the current neutrino mass bound.
	
	This article is arranged as follows. In Section~\ref{sec:theory}, we establish the relevant Lagrangian, set the conventions used in the article, and discuss the approach we follow for computing the renormalisation of the operators involved. In Section~\ref{sec:Running}, we discuss the derivation of the divergences and RGEs and provide these at 1-loop up to order \order{4}. We point out the global structure of the RGEs, and comment on contributions that are larger than expected from naive dimensional analysis. In Section~\ref{sec:Disc}, we derive constraints on LNV parameter space using RGEs and positivity bounds for models where $ \phi^4D^4 $ are missing at tree-level, and perform a case study on the seesaw models. We also discuss bounds induced from the T-parameter to LNVs. Finally, in Section~\ref{sec:Conclusion}, we summarise our findings. We tabulate all the effective operators relevant for our analysis in the Appendix~\ref{app:tables}.

	\section{Theory and methodology}\label{sec:theory}
	\subsection{SMEFT conventions}
	We consider the SMEFT Lagrangian,
	\begin{equation}\label{eq:LagUV}
		\mathcal{L}_\text{SMEFT}=\mathcal{L}_\text{SM} + \left( \alpha_{\ell \phi}\frac{O_{\ell \phi}^{(5)}}{\Lambda} + \text{h.c.} \right) + \sum_i\beta_i\frac{O_i^{(6)}}{\Lambda^2} + \sum_j\omega_j \frac{O_j^{(7)}}{\Lambda^3}+\sum_k c_k\frac{O_k^{(8)}}{\Lambda^4} \, ,
	\end{equation}
	where the expansion in the SMEFT cut-off scale $\Lambda$ is truncated at order~\order{4} and $ \alpha_{\ell \phi} $, $ \beta_i $, $ \omega_j $, and $ c_k $ represent $ d_5, \, d_6, \, d_7, $ and $ d_8 $ Wilson coefficients\footnote{We adopt the convention ``dimension-$n$" as $d_n$ for brevity.}, respectively. The Standard Model Lagrangian is written as:
	\begin{align}\label{eq:SMlag}\nonumber
		\mathcal{L}_\text{SM} = & -\frac{1}{4}G_{\mu\nu}^{A}G^{A\,\mu\nu} -\frac{1}{4}W_{\mu\nu}^{I}W^{I\,\mu\nu} -\frac{1}{4}B_{\mu\nu}B^{\mu\nu}\\\nonumber
		&
		+\overline{q^{m}}\ii\slashed{D}q^{m}
		+\overline{\ell^{m}}\ii\slashed{D}\ell^{m}
		+\overline{u^{m}}\ii\slashed{D}u^{m}
		+\overline{d^{m}}\ii\slashed{D}d^{m}
		+\overline{e^{m}}\ii\slashed{D}e^{m}
		\\
		& +\left(D_{\mu}\phi\right)^{\dagger}\left(D^{\mu}\phi\right)
		+\mu_\phi^{2}|\phi|^{2}-\lambda_\phi |\phi|^{4}
		-\left(
		y^{u}_{mn}\overline{q^{m}}\widetilde{\phi}u^{n}
		+y^{d}_{mn}\overline{q^{m}}\phi d^{n}
		+y^{e}_{mn}\overline{\ell^{m}}\phi e^{n}
		+\text{h.c.}\right) \, .
	\end{align}
	The covariant derivative is defined as:
	\begin{equation}
		D_\mu = \partial_\mu - \ii g_1 Y B_\mu -\ii g_2\frac{\sigma^I}{2} W_\mu^I -\ii g_3\frac{\lambda^A}{2} G_\mu^A\,.
	\end{equation}
	Here, $e$, $u$ and $d$ represent the right-handed leptons, up- and down-type quarks; $\ell$ and $q$ represent the left-handed leptons and quarks; and $B, W,$ and $G$ represent the gauge fields corresponding to $U(1)_Y, \ SU(2)_L$ and $SU(3)_C$, with $g_1, \, g_2$ and $g_3$ as their gauge couplings, respectively. We represent the SM Higgs doublet by $\phi$ $(Y=\frac{1}{2})$, and $\widetilde{\phi}= \ii \sigma^2 \phi^\ast$, where $\sigma$'s denote the Pauli matrices $(I=1,2,3)$. We denote the $U(1)_Y$ hypercharge by $Y$, and $\lambda$'s are the Gell-Mann matrices ($A=1,...,8$).
	
		In the SMEFT expansion, the $d_8$ Wilson coefficients are suppressed by a power of \order{4}. From power counting, we deduce that the running of $d_8$ at this order is triggered by the insertions of (I) one $ d_8 $ \cite{DasBakshi:2022mwk}, (II) two $ d_6 $ \cite{Chala:2021pll}, (III) one $ d_5 $ and one $ d_7 $, (IV) two $d_5$ and one $d_6$ or (V) four $d_5$ Wilson coefficients. The Weinberg operator is the only operator at $ d_5 $ \cite{PhysRevLett.43.1566}. We use physical bases for the insertions of $d_6$ \cite{Grzadkowski:2010es} and $d_7$ \cite{Lehman:2014jma}. However, the method employed to compute the divergences requires Green's bases for $d_6$ \cite{Gherardi:2020det} and $d_8$ \cite{Chala:2021cgt} \footnote[3]{Other choices for SMEFT $d_7$ and $d_8$ operators basis are available at \cite{Liao:2016hru,Li:2020gnx,Ren:2022tvi}.}.
	
	We perform the renormalisation of the $d_8$ Wilson coefficients using an off-shell scheme \cite{Chala:2021cgt}. The rationale of this method consists in considering only diagrams that are 1-particle irreducible (1PI) to generate the divergences of the Wilson coefficients (including those of non-physical operators). In general, we do this by extending the physical basis with a set of redundant operators spanning all the possible interactions.
	
	We compute the off-shell 1PI diagrams to generate amplitudes, which are then captured in the Wilson coefficients of the Green's basis operators. The redundant operators can be reduced to the physical subset by applying field redefinitions or the Equations of Motion (EoMs). Consequently, we find that the contribution of the redundant operators amounts to a shift in the definition of the Wilson coefficients of the physical operators:  
	\begin{equation}\label{eq:wcshiftgeneric}
		\frac{c_i^\text{phys}}{\Lambda^n}\rightarrow \frac{c_i^\text{phys}}{\Lambda^n} + \frac{1}{\Lambda^n}\sum_k b_k c_k^\text{red} \,,
	\end{equation}
	where, $b_k$ usually contains powers of the SM couplings but can also enclose SMEFT coefficients of lower dimension provided the mass dimension of each term in the shift is the same. Combinations of more than one SMEFT WC appear when the EoMs for some field are applied at a greater order in $\Lambda$. For $d_8$ WCs, this means that the shift of the physical coefficients contain linear combinations of redundant $d_8$ WCs but also pairs of redundant $d_6$ WCs. In principle we could also find contributions of $d_5$ and $d_7$ WCs but applying the EoMs to these operators leads only to fermionic operators, which are out of the scope of this paper. Hence, the only redundant degrees of freedom we need to consider are those contained in the Green's bases for $d_6$ and $d_8$.
	
	The computation is feasible without introducing redundant operators, but it would imply dealing with all connected diagrams. In contrast, the off-shell approach just adds an extra systematical step translating the WCs from the Green's basis to the physical basis, and it is already worked out for bosonic interactions in Ref.~\cite{Chala:2021cgt}. In practice, the contribution from connected diagrams in the on-shell approach is indirectly taken care of when the WC shift is applied since it contains the information given by the EoMs.
	
	\subsection{Organizing the calculation}\label{sec:sortingcontribs}
	
	The main goal of this paper is to compute the contributions of all LNV operators to the RGEs of the $d_8$ bosonic Wilson coefficients and to order~\order{4}. At first, one may think of the many possible different topologies generated by combinations of $d_5$, $d_6$ and $d_7$ operators and SM couplings. Considering that the great number of operators (1+20+63) can lead to several valid diagrams, it seems inevitable that the amplitudes for some processes get contributions from many operators. Restricting to bosonic processes removes a significant number of these, and -after a more exhaustive analysis- we show more valid contributions that actually vanish, thus remarkably reducing the effort needed for the whole computation. In Table~\ref{tab:BosContribs}, we schematically show the variety of possibilities and the diagrams we are left with. Throughout the rest of this section, we elaborate on the reasoning for which most of these contributions vanish.
	\begin{table}[h]
		\begin{center}
			\renewcommand{\arraystretch}{1.5}
			\begin{tabular}{|c|c|c|c|}
				\hline
				$1/\Lambda^4$ & $d_5^4$ & $d_5^2\times d_6$ & $d_5\times d_7$ \\
				\hline \hline
				8-Higgs & \ref{fig:d5^4} & - & - \\
				\hline
				6-Higgs & - & \ref{fig:d5^2_psi2phi2D} & \ref{fig:d5_psi2phi4}, \ref{fig:Y2_d5_psi2phi2D2}, \ref{fig:Y_d5_psi2phi3D} \\
				\hline
				4-Higgs & - & - & \ref{fig:d5_psi2phi2D2}, \ref{fig:g_d5_psi2phi2D2}, \ref{fig:g2_d5_psi2phi2D2}, \ref{fig:g_d5_psi2phi2X}\\
				\hline
				2-Higgs & - & - & - \\
				\hline
				0-Higgs & - & - & - \\
				\hline 
			\end{tabular}
		\end{center}
		\caption{\label{tab:BosContribs} Relevant contributions for the renormalisation of bosonic operators to order \order{4}. Rows represent the classes of $d_8$ bosonic operators, grouped by the number of Higgs they contain. Columns represent the possibilities for the insertion of LNV operators. A number means the contribution is sizeable and considered in this paper in the figure corresponding to that number. A hyphen (-) means the contribution vanishes or is out of the scope of this paper. See the text for more details.}
	\end{table}

	As we can see in the Lagrangian Eq.~\ref{eq:LagUV}, the WCs are suppressed by a power of the energy scale $\Lambda$ related to the mass dimension of the operator it refers to. With this Lagrangian and to order \order{4} the following contributions need to be considered: $(d_5^4)$, $(d_5^2\times d_6)$ and $(d_5 \times d_7)$. The divergences of the bosonic $d_8$ operators with $d_6^2$ and the self-contribution of $d_8$ are available in \cite{Chala:2021pll} and \cite{DasBakshi:2022mwk}, respectively, so we do not consider them anymore. What remains is precisely the aim of this work: contributions including at least one LNV operator. Before discussing each set of insertions individually, there are some global remarks that apply to all cases.
	
	A priori, we deal with insertions of $d_5$, $d_6$ and $d_7$ combined as $d_5^4$, $d_5^2\times d_6$ and $d_5 \times d_7$ to make 1PI diagrams of processes that only include bosons as external fields, as we renormalise the bosonic $d_8$ operators. That would be the maximum number of contributions we would have to compute, but now we argue that most of the calculations vanish, and the nonzero combinations are but a small subset of the total amount of possibilities.
	
	First of all, notice that the field content of the loops in the diagrams can only consist of fermions, and not bosons. It follows from the fact that all of the LNV operators inserted include one or two pairs of fermion fields (a bispinor and its conjugate for each pair), which generate a vertex with two or four fermionic legs. Since the operators we strive to renormalise do not contain fermions, the fermionic legs cannot be external, thus forcing them to be closed in a loop. This directly affects the renormalisation operators of mass dimension lower than eight (See subsection \ref{sec:belowd8}).

	\begin{figure}
		\begin{center}
			\subfloat[][]{\includegraphics[scale=0.06]{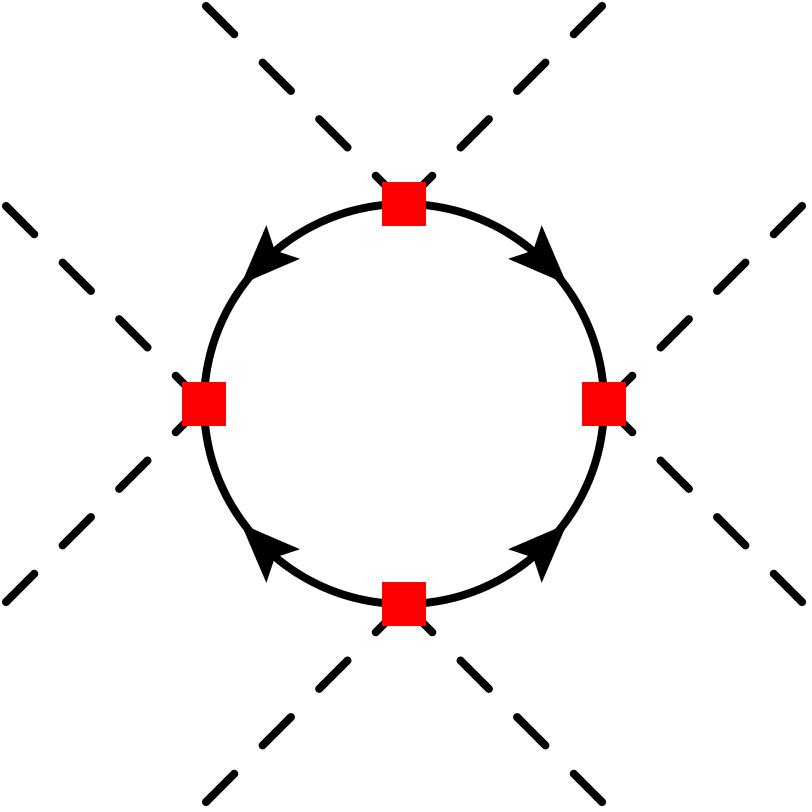}\label{fig:d5^4}} \qquad
			\subfloat[][]{\includegraphics[scale=0.06]{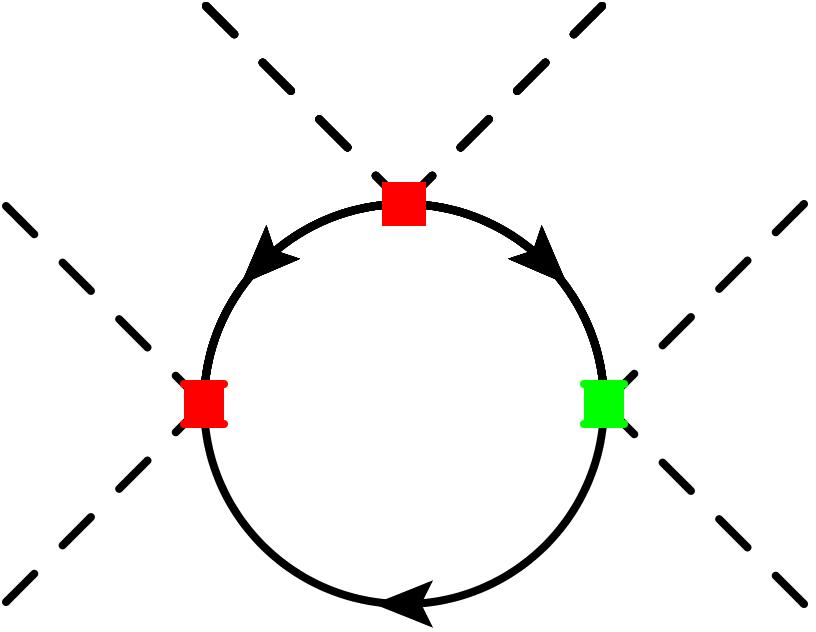}\label{fig:d5^2_psi2phi2D}} \qquad
			\subfloat[][]{\includegraphics[scale=0.06]{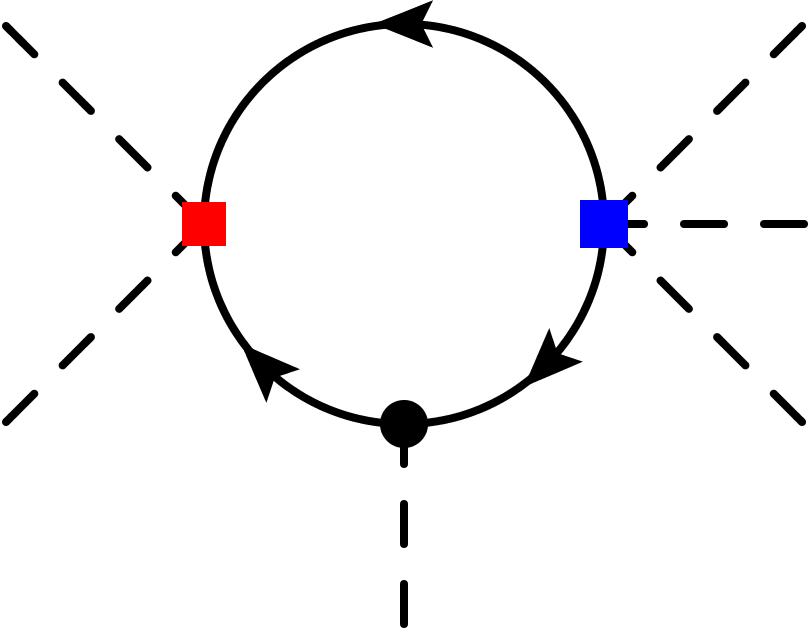}\label{fig:Y_d5_psi2phi3D}}\\
			\subfloat[][]{\includegraphics[scale=0.06]{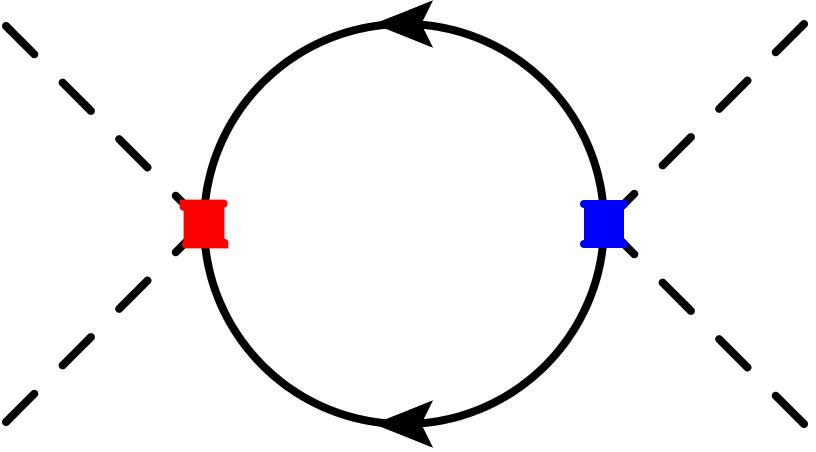}\label{fig:d5_psi2phi2D2}}\qquad 
			\subfloat[][]{\includegraphics[scale=0.06]{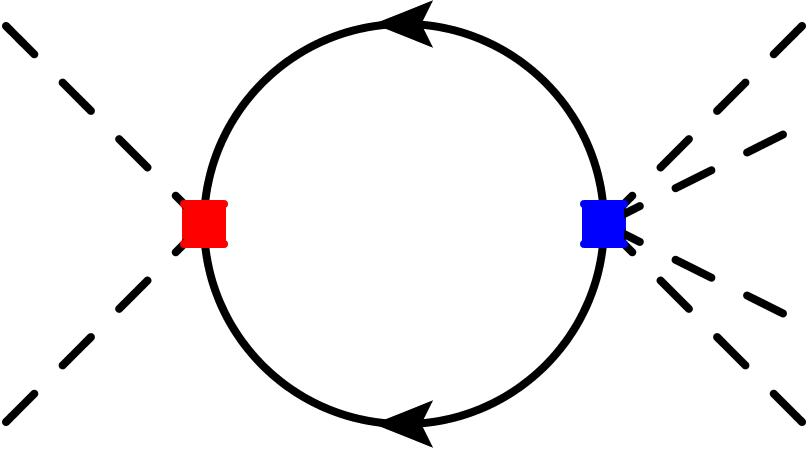}\label{fig:d5_psi2phi4}}\qquad 
			\subfloat[][]{\includegraphics[scale=0.06]{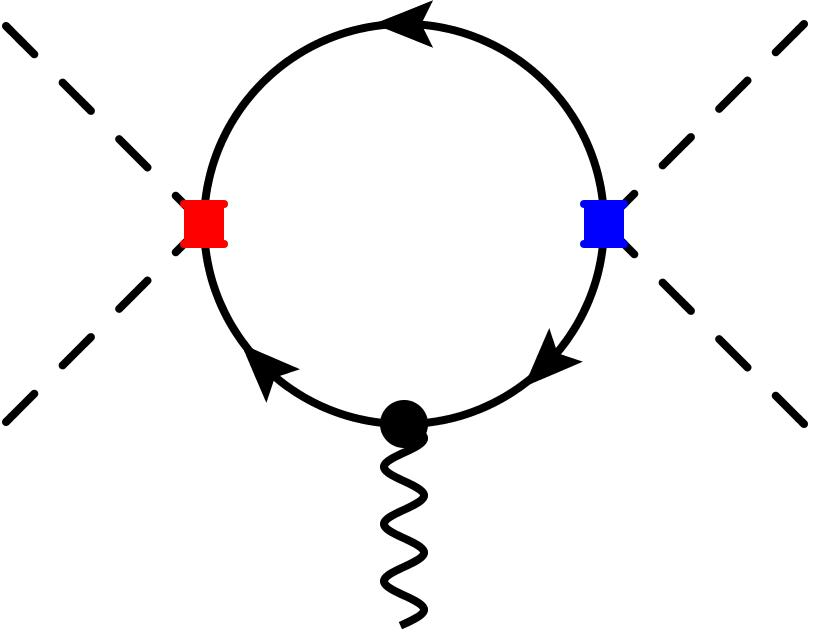}\label{fig:g_d5_psi2phi2D2}}\\
			\subfloat[][]{\includegraphics[scale=0.06]{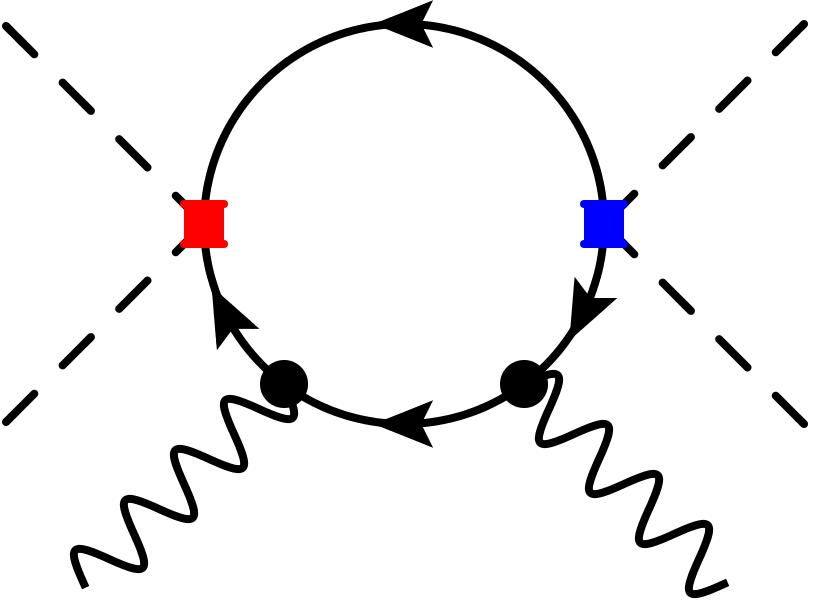}\label{fig:g2_d5_psi2phi2D2}}\qquad 
			\subfloat[][]{\includegraphics[scale=0.06]{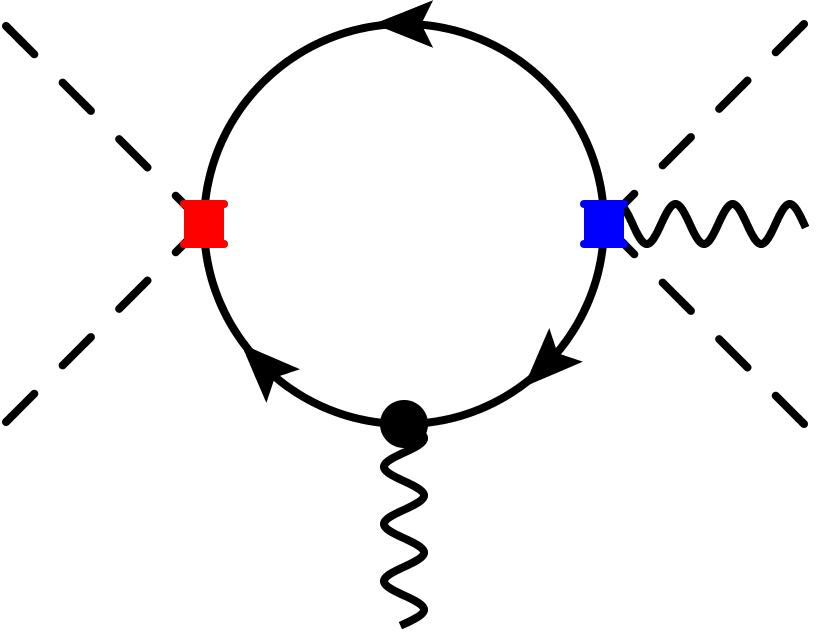}\label{fig:g_d5_psi2phi2X}} \qquad
			\subfloat[][]{\includegraphics[scale=0.06]{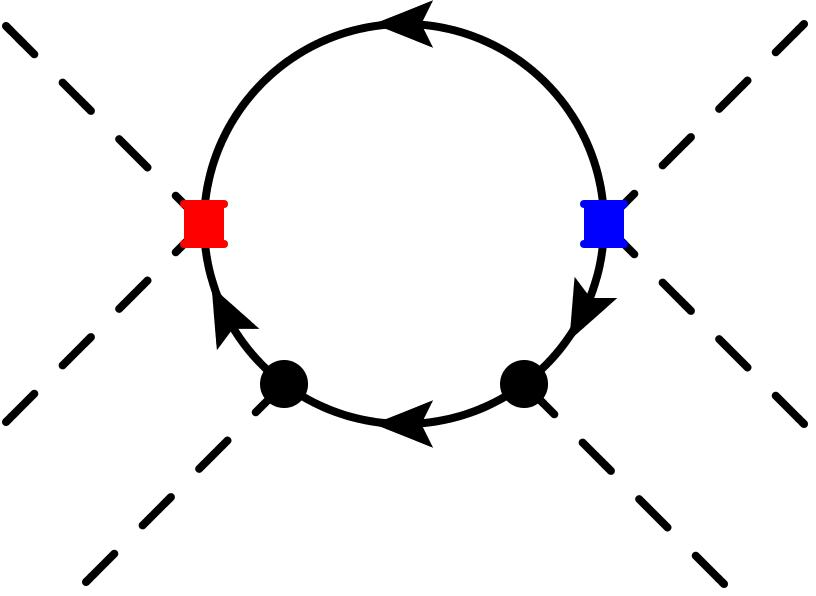}\label{fig:Y2_d5_psi2phi2D2}}%
		\end{center}
		\caption{\label{fig:LNVcontrib} Contributions from all possible insertions of LNV operators to order \order{4}. Vertices in \textcolor{red}{red}, \textcolor{green}{green} and \textcolor{blue}{blue} represent insertions of a $d_5$, $d_6$, and $d_7$ operators, respectively. Dashed, solid, and wavy lines represent scalar, fermion, and gauge boson propagators/legs, respectively.}
	\end{figure}
	
	\begin{figure}
		\begin{center}
			\subfloat[][]{\includegraphics[scale=0.06]{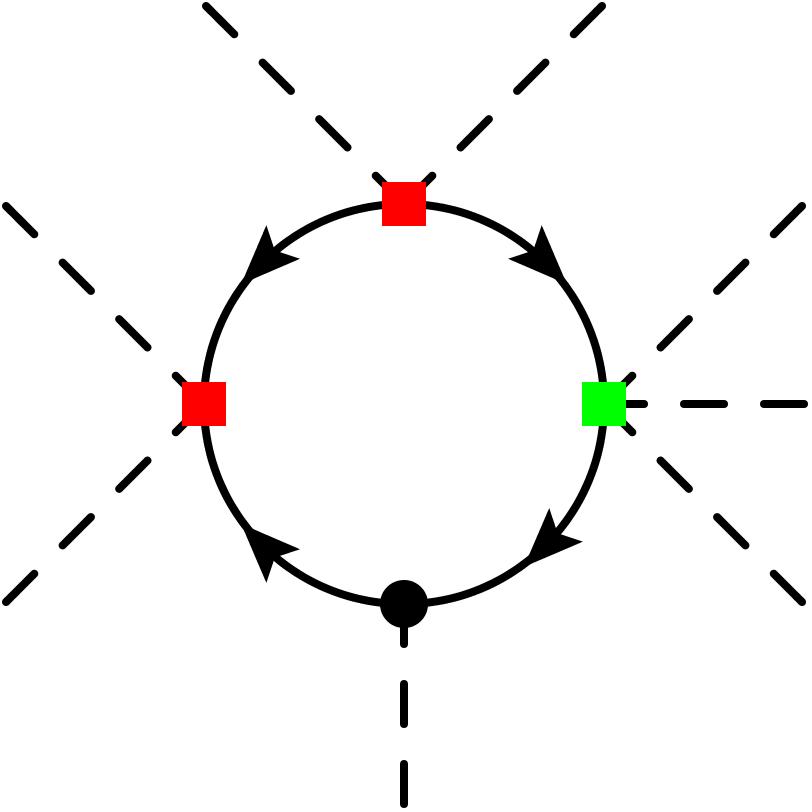} \label{fig:Y_d5^2_psi2phi3}} \qquad 
			\subfloat[][]{\includegraphics[scale=0.06]{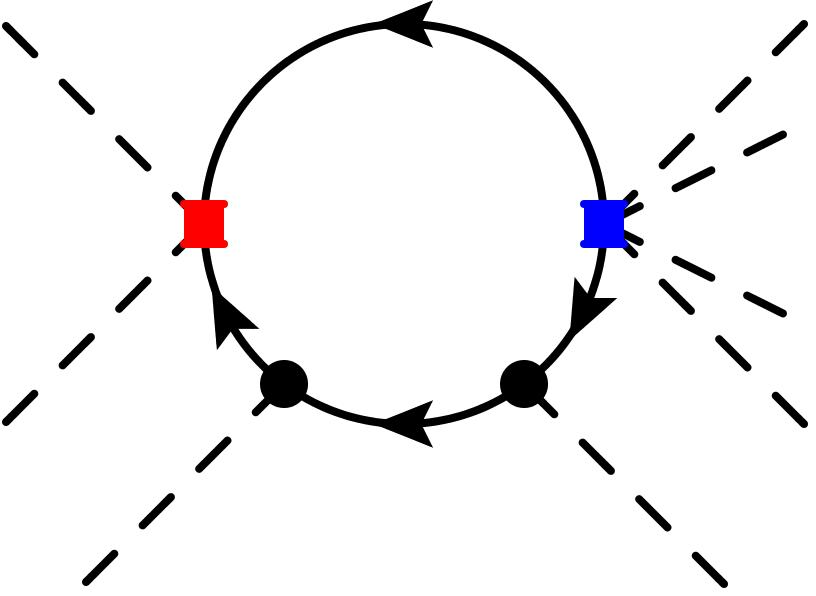} \label{fig:2Y_d5_psi2phi4}} \qquad
			\subfloat[][]{\includegraphics[scale=0.06]{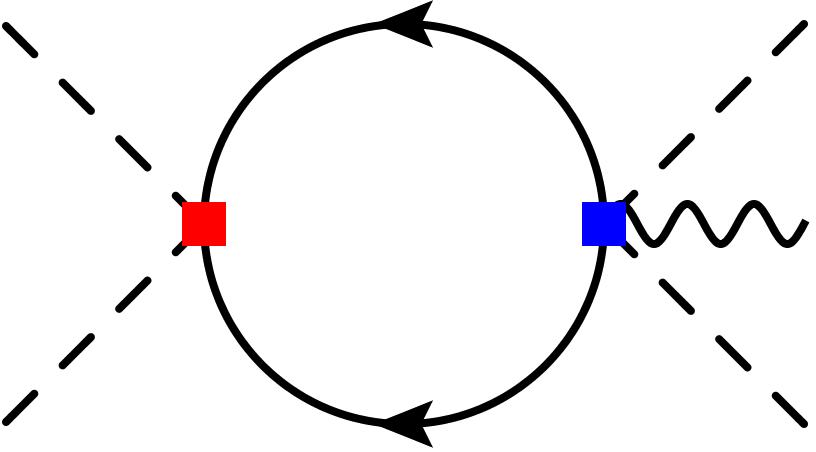}\label{fig:d5_psi2phi2X}} \\
			\subfloat[][]{\includegraphics[scale=0.06]{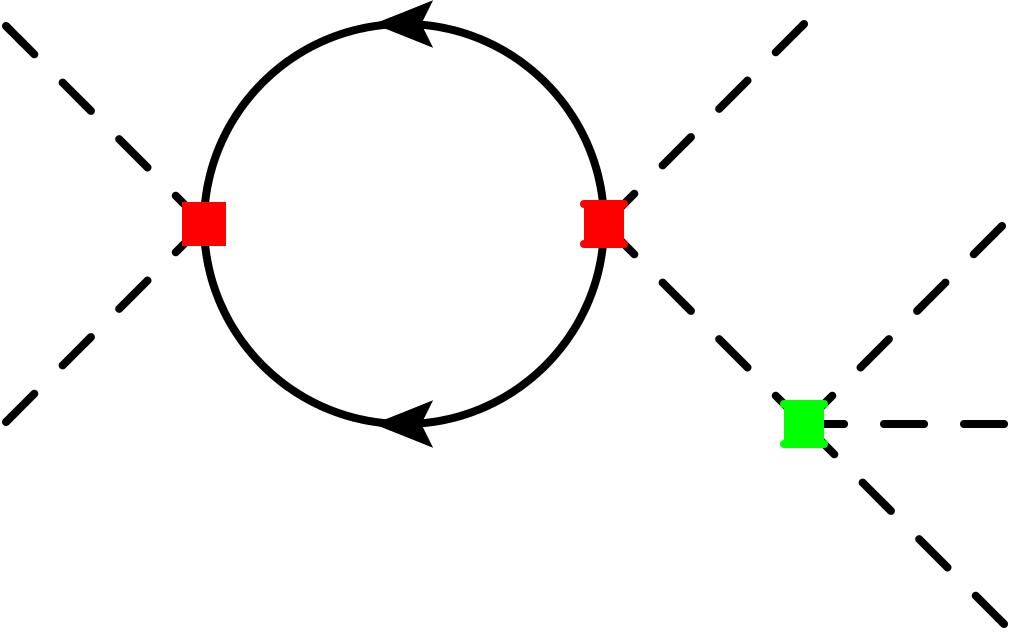}\label{fig:phi4D2onshell}} \qquad
			\subfloat[][]{\includegraphics[scale=0.06]{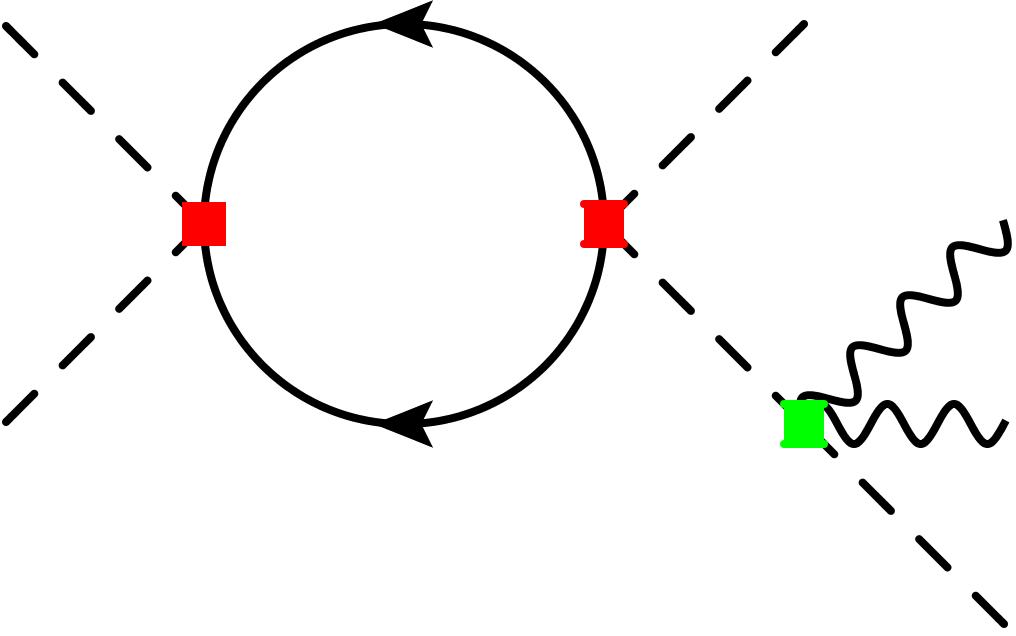}\label{fig:X2phi2onshell}} \qquad 
			\subfloat[][]{\includegraphics[scale=0.06]{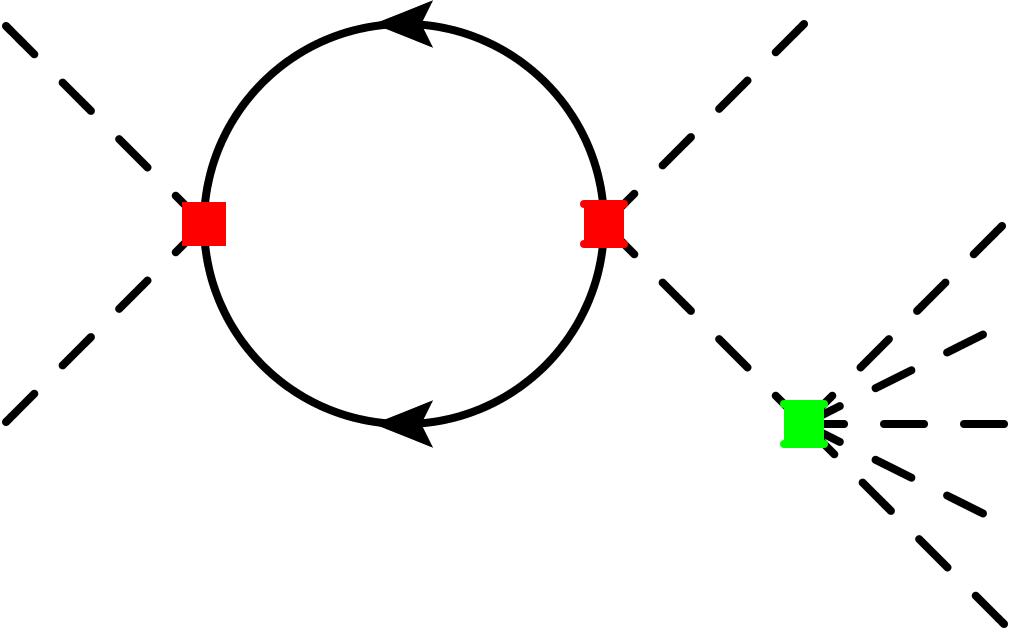} \label{fig:phi6onshell}} %
		\end{center}
		\caption{\label{fig:zerocontrib} Relevant diagrams with insertions of LNV operators. Diagrams \ref{fig:Y_d5^2_psi2phi3} and \ref{fig:2Y_d5_psi2phi4} vanish due to symmetries. \ref{fig:phi4D2onshell} is included in the Wilson coefficient shift. \ref{fig:X2phi2onshell} and \ref{fig:phi6onshell} are also included in the Wilson coefficient shift, but their overall contribution vanishes after computing all the divergences. Vertices in \textcolor{red}{red}, \textcolor{green}{green} and \textcolor{blue}{blue} represent insertions of a $d_5$, $d_6$, or $d_7$ operators, respectively.}
	\end{figure}
	
	The point made above can be exploited from a different perspective. Since fermion lines must be closed, then the contribution of a four-fermion operator to a certain bosonic WC would necessarily lead to a 2-loop diagram, thus being out of the scope of this analysis. Therefore, we only need to consider operators with less than four fermions at the loop. In particular, this means baryon and lepton number violating operators from the $d_6$ set cannot renormalise the bosonic operators at 1-loop order.
	
	We do not need to consider all operators for insertions, even if they meet the requirements above. Among the different classes of $d_6$ operators  we find some of them that are not tree-level generated in the UV completions of SMEFT \cite{Craig:2019wmo}. This means that any diagram including these operators would effectively be loop-suppressed with respect to those that include only tree-level generated operators. The loop-generated operators would lead to a process of greater order in loop expansion, which we ignore for this paper.
	
	One last general remark about the divergent diagrams is in order: there are $d_8$  operators containing more than six fields in their definition (i.e., $\mathcal{O}_{\phi^8}$, $\mathcal{O}^{(1,2,3,4)}_{\phi^6}$). For these to be renormalised, we need to find a diagram with as many legs as fields. It is somewhat easy to reach this number with certain insertions of operators; for example, with the Weinberg operator and a $d_7$ operator from class $\psi^2\phi^4$ we get six Higgs as external fields (diagram \ref{fig:d5_psi2phi4}). One can also try reaching six or eight external legs using SM couplings like Yukawas or gauge boson couplings. However, since we are only considering 1PI diagrams, any vertex we add increases the number of propagators in the loop, which can render the diagram finite. Thus, the insertion of operators is capped.
	
	Regarding the $d_8$ Wilson coefficients, not all 89+86 physical and redundant bosonic operators are renormalised. Recalling the Weinberg operator 
	\begin{equation}
		\mathcal{O}_{\ell \phi}^{(5)}=\epsilon_{ij}\epsilon_{mn}\left(\ell^iC\ell^m\right)\phi^j\phi^n,
	\end{equation}
	we see that two Higgs fields appear as external legs since the loop is formed by the fermion lines. Thus, the insertion of $\alpha_{\ell \phi}$ implies that the final diagram will have at least two Higgs as external legs, and we ignore all those $d_8$ operators that do not include Higgs fields. Furthermore, considering that for $d_5^4$ and $d_5^2\times d_6$ there are two or more insertions of the Weinberg operator, then diagrams with these insertions do not contribute to operators with less than four Higgs fields. It so happens that all of the $d_7$ operators meeting the previous properties have at least two Higgs fields as well, so the diagrams for $d_5 \times d_7$ have at least four Higgs external legs too. In conclusion, there are no contributions from LNV insertions to the RGEs of operators with less than four Higgs fields, including all of the loop-generated bosonic $d_8$ operators.
	
	Finally, we argue why most operators featuring gauge bosons can be neglected too. All instances containing a gluon field are not renormalised since there are no gluonic LNV operators. The renormalisation of operators with $B$ bosons is also heavily restricted due to charge conservation or other symmetries.
	
	Summing up,
	\begin{itemize}
		\item For all diagrams with insertions of LNV, there is only one loop always formed by fermions.
		\item Only 2-fermion operators are inserted. In particular, this means the $d_6$ LNV and baryon number violating operators cannot renormalise the $d_8$ bosonic operators.
		\item Only tree-level generated $ d_6 $ operators are inserted.
		\item There is a limit to the number of Yukawa and gauge couplings that can be inserted in a diagram before it is no longer divergent.
		\item Operators with less than four Higgs fields do not get any contribution to the RGEs. This includes class $X^2\phi^2D^2$.
	
	\end{itemize}
	
	These points greatly reduce the scenarios of possible insertions to the computations needed thus simplifying their difficulty. Nevertheless, we can take this analysis further by focusing on each combination.
	\subsubsection{Insertions of $d_5 \times d_7 $ \label{sec:sortd5d7}}
	
	Let us study a particular case of $d_5 \times d_7$. If we consider the vertex generated by the $d_7$ Weinberg operator,
	\begin{equation}
		\mathcal{O}_{\ell \phi}^{(7)}=\epsilon_{ij}\epsilon_{mn}\left(\ell^iC\ell^m\right)\phi^j\phi^n\left(\phi^\dagger \phi\right),
	\end{equation} 
	we could link the $d_5$ operator vertex in several different ways. However, since we are interested in the bosonic contributions, we must close all fermion lines. There is only one way to do this for the purpose of renormalisation at 1-loop, which is joining the fermion lines of both operators so that a loop is formed. Thus, we discard contributions to operators with less than six Higgs legs from diagram~\ref{fig:d5_psi2phi4}.
	
	There are more $d_7$ operators, which could lead to different contributions, but some vanish. This is the case of Figure~\ref{fig:2Y_d5_psi2phi4}, which represents a topology where we could have the insertion of $\mathcal{O}_{\ell \phi}^{(5)}$ and $\left(\mathcal{O}_{\ell \phi}^{(7)}\right)^\dagger$ or $\left(\mathcal{O}_{\ell \phi}^{(5)}\right)^\dagger$ and $\mathcal{O}_{\ell \phi}^{(7)}$. However, we find that the total amplitude vanishes due to the cancellation among diagrams which differ in permutations of internal loop propagators. The presence of the Yukawas is essential for this cancellation; in particular, diagram \ref{fig:d5_psi2phi4} looks similar to \ref{fig:2Y_d5_psi2phi4} but does not vanish for those insertions. We do not see such cancellations when the $d_7$ operator inserted is from class $\psi^2\phi^2 D^2$ since there is no symmetry in their definitions, so diagrams like \ref{fig:Y2_d5_psi2phi2D2} and \ref{fig:g2_d5_psi2phi2D2} are nonzero. 
	
	\ref{fig:2Y_d5_psi2phi4} is actually the only possible divergence of $d_5\times d_7$ with eight Higgs in the external legs. Knowing its contribution vanishes implies no direct contribution from $d_5\times d_7$ appear in the divergences of the 8-Higgs operator while we can see direct contributions in operators with less number of Higgs.
	
Regarding the renormalisation of operatos with gauge bosons, the only LNV operators containing gauge fields are $\mathcal{O}_{\ell \phi B}$ and $\mathcal{O}_{\ell \phi W}$ which, by definition, are skew-symmetric to flavor transposition, while $\mathcal{O}^{(5)}_{\ell \phi}$ is symmetric:
\begin{equation}
	\left(\mathcal{O}_{\ell \phi B,\,W}\right)_{pq}=-\left(\mathcal{O}_{\ell \phi B,\,W}\right)_{qp} \, , \qquad \left(\mathcal{O}^{(5)}_{\ell \phi}\right)_{pq}=\left(\mathcal{O}^{(5)}_{\ell \phi}\right)_{qp}\,.
\end{equation}
It follows from this fact, that the amplitude of diagrams like \ref{fig:d5_psi2phi2X} will trivially vanish since they are proportional to the trace of these two operators. For the same reason, diagram \ref{fig:g_d5_psi2phi2X} will also lead to null amplitudes unless the two gauge bosons in the process are $W$ bosons. In that case, the commutation of Pauli matrices can lead to non-vanishing terms.
	
	\subsubsection{Insertions of $ d_5 \times d_5 \times d_6 $}
	We could use a similar reasoning for the insertion of two Weinberg operators and one $d_6$ operator. It is clear by the points argued before that both $d_5$ operators must be in the loop. A $d_6$ operator will also be in the loop if it has fermionic lines, forming the diagrams in Figure~\ref{fig:d5^2_psi2phi2D}. Note how there are no physical tree-level generated $d_6$ two-fermion operators without at least two Higgs fields, implying we cannot draw a diagram with less than 6-Higgs as external legs, and so, there do not be any direct contribution to operators with 4-Higgs legs ($ i.e. $ classes $\phi^4 D^4$, $X\phi^4 D^2$ and $X^2 \phi^4$). 
	
	However, we can also think of insertions of bosonic $d_6$ operators. In this case, precisely due to the lack of fermions, the vertex cannot be part of the loop. Instead, it appears in an external leg, thus forming a 1-particle-reducible diagram. We have chosen not to consider these in the off-shell approach (as argued in Section~\ref{sec:theory}), so we discard diagrams like \ref{fig:phi4D2onshell}, \ref{fig:X2phi2onshell} or \ref{fig:phi6onshell}. Instead, we use the divergences of the generated redundant operators to order~\order{2}, which yield \order{4} terms after the WC shift.

	Once again, there is a remark about the contribution to the 8-Higgs operator, $\mathcal{O}_{\phi^8}$. For the 1PI amplitudes, Figure~\ref{fig:Y_d5^2_psi2phi3} shows the diagram that would need to be considered. Just as \ref{fig:2Y_d5_psi2phi4}, the total amplitude for the insertion of $\mathcal{O}_{\ell \phi}^{(5)}$ and $\mathcal{O}_{\ell \phi}$ vanishes due to mutual cancellation among the diagrams.

	\subsubsection{Insertions of $ d_5 \times d_5 \times d_5 \times d_5 $}
	Finally, we arrive at the contribution of four Weinberg operators inserted in one loop. Since fermion lines must be closed, their WCs will appear contracted in the same trace, resulting in eight Higgs as external legs (Figure~\ref{fig:d5^4}). This clearly contributes to the 8-Higgs $d_8$ operator
	\begin{equation}
		\mathcal{O}_{\phi^8}=\left(\phi^\dagger \phi\right)^4\,.
	\end{equation}
	The amplitude is proportional to a trace of the four WCs. The eight legs of the diagram are the maximum number allowed for an amplitude of order~\order{4}. Since there is only the Weinberg operator inserted several times, the remaining freedom resides in the fermions of the loop. Summing all the combinations yields the only contribution from this configuration.
	
	As argued in the previous sections, the $\phi^8$ class of $d_8$ operators receives no other contribution from 1PI diagrams, meaning the only term in Eq.~\ref{eq:divergencesd8} accounts for all the divergences of $\mathcal{O}_{\phi^8}$ coming from the LNV operators.
	
	\subsubsection{Renormalisation below $d_8$}\label{sec:belowd8}
	Here, we comment on the renormalisation of operators with lower mass dimension. 
	Since we are only considering contributions to bosonic operators, it is clear that there must be an even number of LNV operators inserted to get a non-LNV contribution, which implies that the Wilson coefficients could only be renormalised to order \order{2} by two insertions of the Weinberg operator: $d_5^2$. The bosonic operator's coefficients of the renormalisable SM do not get a contribution from these, and for the $d_6$ bosonic operators are computed in Ref.~\cite{Davidson:2018zuo}.
	
	About the renormalisation to order \order{4} of operators below $d_8$, we know the divergences carry a factor of \order{4} in the amplitude compensated by another mass scale to get the correct mass dimension below eight for the operators. The only mass scale in the Lagrangian Eq.~\ref{eq:LagUV} is $\mu_\phi$, meaning that a loop of Higgs fields would need to appear in the diagram. We have already established in Section~\ref{sec:sortingcontribs} that the loop for the bosonic contributions needs to be formed by fermions, thus no contribution to order \order{4} is added \textit{directly} by $d_5^4$, $d_5^2\times d_6$ and $d_5 \times d_7$ to the RGEs of Wilson coefficients from bosonic operators with mass dimension below 8. Nonetheless, this does not necessarily imply that there is no contribution. In the off-shell approach, these contribution is encoded in the non-physical operators from the Green's basis. Once the divergences have been computed, one must shift the physical coefficients (as discussed in Ref.~\cite{Chala:2021cgt}), this is a consequence of applying the Equations of Motion. The shift causes a mixing of operators of different classes with coupling factors which, in fact, give a nonzero indirect contribution to the RGE of some operators even if all their divergent 1PI diagrams are null.
	
	\section{The RGEs}\label{sec:Running}
	In this section, we present the RGEs of the $d_8$ Wilson coefficients up to order \order{4} and discuss the derivation in detail. The main results have been uploaded to a public \href{https://github.com/SMEFT-Dimension8-RGEs/Notebooks}{Mathematica notebook}.
	
	In Section~\ref{sec:divergences}, we show the divergences we use to get the RGEs in \ref{sec:RGEslist}, where we also tabulated the greatest contributions to the anomalous dimension matrix. The process is simple, but we explain it here for a complete analysis.
	
	In general, we shall have three sources of LNV contributions to the anomalous dimension matrix to order \order{4}:
	\begin{equation}\label{eq:rgedefinition}
		16\pi^2\frac{\rm{d}}{\rm{d}\ln{\mu}}c_k(\mu) := \dot{c}_k=\gamma_{k}\alpha_{\ell \phi}\alpha_{\ell \phi}\alpha_{\ell \phi}\alpha_{\ell \phi} + \gamma_{ik}\alpha_{\ell \phi}\alpha_{\ell \phi}\beta_i + \gamma_{jk}\alpha_{\ell \phi}\omega_j\,,
	\end{equation}
	where hereafter $c_k$ are Wilson coefficients of $d_8$ bosonic operators (shown in Table~\ref{tab:dim8ops1}), but any other coefficient of dimension lower than eight can be either bosonic or fermionic (see Table~\ref{tab:op67}). Note this only affects $\beta_i$, which are the $d_6$ Wilson coefficients -- the $d_5$ ($\alpha_{\ell \phi}$) and $d_7$ ($\omega_j$) only include fermionic operators.
	
	Following the steps in section~\ref{sec:theory}, the divergences are obtained. Now, they are applied to the formula for the computation of the RGEs up to 1-loop order:
	\begin{equation}\label{eq:rgedef}
		\dot{c}_i=-c_i\sum_j n_j x_j \frac{\partial}{\partial x_j}\left(\frac{ \tilde{c}_i}{c_i} \right),
	\end{equation}
	where $n_j$ is the tree-level anomalous dimension\footnote{For an operator with $n_\psi$ fermions, $n_X$ field strength tensors, $n_\phi$ Higgs fields then $n_j=n_\psi+n_X+n_\phi-2$ in $D=4-2\epsilon$.}, and $x_j$ are all the Wilson coefficients (running or not) that enter the Lagrangian. $c_i$ are the Wilson coefficients that are renormalised, whereas, $\tilde{c}_i$ are the computed divergences. They should also include contributions from the field redefinition needed to normalise the Higgs kinetic term canonically, as well as the Wave Function Renormalisation (WFR) for gauge bosons; however, these only affect self-contributions to the RGEs, which we do not consider here.
	
	For an exhaustive computation of the anomalous dimension matrix to order~\order{4}, we must also consider the divergences of $d_6$ operators to order~\order{2} by LNV operators. The insertions can only be of two $d_5$ operators. Nonetheless, we have attached these results in Section~\ref{sec:divergences} for completeness.
	
	\subsection{Divergences}\label{sec:divergences}
	
	The calculation of the divergences is the key step to determine the running of the Wilson coefficients. We have already shown in Section \ref{sec:theory} the 1PI diagrams whose amplitudes we need to compute for renormalisation. These divergent amplitudes are matched with those of the corresponding tree-level processes with $d_8$ vertices.
	
	In order to do this, both the SMEFT and the $d_8$ effective Lagrangians are implemented in \texttt{FeynRules} \cite{Alloul:2013bka} models, and the amplitudes are computed with the help of \texttt{FormCalc} \cite{Hahn:1998yk} and \texttt{FeynArts} \cite{Hahn:2000kx}. The amplitudes are a sum of expressions that depend on the couplings and kinematic invariants (\textit{i.e.} a set of algebraically independent contractions of momenta and polarisation vectors). Equating the amplitudes for each process, we get a set of as many equations as unique kinematic invariants. The solutions of these equations are the divergences for the $d_8$ Wilson coefficients of the Green's basis operators, which we present here.\footnote{The divergences of classes $\phi^8$ and $\phi^6 D^2$ are cross-checked using \texttt{matchmakereft} \cite{Carmona:2021xtq}.}
	
	\begin{align} \label{eq:divergencesd8}
		\tilde{c}_{\phi^8} &\supset - \frac{1}{2\pi^2\epsilon} \text{Tr} \left[\alpha_{\ell \phi}\alpha_{\ell \phi}^\dagger \alpha_{\ell \phi}\alpha_{\ell \phi}^\dagger\right],\\[1em]
		\tilde{c}_{\phi^6 D^2}^{(1)} &\supset \frac{1}{2\pi^2\epsilon} \text{Tr} \left[-3\beta_{\phi \ell}^{(1)}\alpha_{\ell \phi}^\dagger \alpha_{\ell \phi} + 4\beta_{\phi \ell}^{(3)} \alpha_{\ell \phi}^\dagger \alpha_{\ell \phi} \right] + \frac{1}{4\pi^2\epsilon}\text{Im}\left(\text{Tr} \left[\alpha_{\ell \phi}y^e\omega_{\ell \phi D e}^* \right] \right)\nonumber \\
		&+ \frac{1}{2\pi^2\epsilon}\text{Re}\left(\text{Tr} \left[\alpha_{\ell \phi} \omega_{\ell \phi}^\dagger\right] \right) - \frac{1}{16\pi^2\epsilon}\text{Re}\left(\text{Tr} \left[\alpha_{\ell \phi}^\dagger \left(y^e\right)^* \left(y^e\right) \omega_{\ell \phi D}^{(2)}  \right] \right) \nonumber \\
		&- \frac{1}{16\pi^2\epsilon}\text{Re}\left(\text{Tr} \left[\alpha_{\ell \phi}^\dagger  \left(y^e\right)^* \left(y^e\right) \left(\omega_{\ell \phi D}^{(1)} + \left(\omega_{\ell \phi D}^{(1)}\right)^T \right)\right] \right),\\[1em]
		\tilde{c}_{\phi^6 D^2}^{(2)} &\supset \frac{1}{2\pi^2\epsilon} \text{Tr} \left[-2\beta_{\phi \ell}^{(1)}\alpha_{\ell \phi}^\dagger \alpha_{\ell \phi} + \beta_{\phi \ell}^{(3)}\alpha_{\ell \phi}^\dagger \alpha_{\ell \phi} \right] - \frac{1}{4\pi^2\epsilon}\text{Im}\left(\text{Tr} \left[\alpha_{\ell \phi} y^e \omega_{\ell \phi D e}^* \right] \right) \nonumber \\
		& + \frac{1}{4\pi^2\epsilon}\text{Re}\left(\text{Tr} \left[\alpha_{\ell \phi} \omega_{\ell \phi}^\dagger\right] \right) - \frac{1}{16\pi^2\epsilon}\text{Re}\left(\text{Tr} \left[\alpha_{\ell \phi}^\dagger \left(y^e\right)^* \left(y^e\right) \omega_{\ell \phi D}^{(2)}  \right] \right)\nonumber\\
		&- \frac{1}{16\pi^2\epsilon}\text{Re}\left(\text{Tr} \left[\alpha_{\ell \phi}^\dagger  \left(y^e\right)^* \left(y^e\right) \left(\omega_{\ell \phi D}^{(1)} + \left(\omega_{\ell \phi D}^{(1)}\right)^T \right)\right] \right),\\[1em]
		\tilde{c}_{\phi^6 D^2}^{(3)} &\supset \frac{1}{4\pi^2\epsilon} \text{Tr} \left[-\beta_{\phi \ell}^{(1)}\alpha_{\ell \phi}^\dagger \alpha_{\ell \phi} + \beta_{\phi \ell}^{(3)} \alpha_{\ell \phi}^\dagger \alpha_{\ell \phi}\right] - \frac{1}{8\pi^2\epsilon}\text{Re}\left(\text{Tr} \left[\alpha_{\ell \phi} \omega_{\ell \phi}^\dagger\right] \right),\\[1em]
		\tilde{c}_{\phi^6 D^2}^{(4)} &\supset \frac{1}{8\pi^2\epsilon}\text{Im}\left(\text{Tr} \left[\alpha_{\ell \phi} \omega_{\ell \phi}^\dagger\right] \right), \\[1em]
		\tilde{c}_{\phi^4 D^4}^{(2)} &\supset \frac{1}{4\pi^2\epsilon}\text{Re}\left(\text{Tr} \left[\alpha_{\ell \phi}^\dagger \omega_{\ell \phi D}^{(2)}\right] \right), \\[1em]
		\tilde{c}_{\phi^4 D^4}^{(10)} &\supset \frac{1}{8\pi^2\epsilon}\text{Re}\left(\text{Tr} \left[\alpha_{\ell \phi}^\dagger \omega_{\ell \phi D}^{(2)}\right] \right), \\[1em]
		\tilde{c}_{\phi^4 D^4}^{(11)} &\supset \frac{1}{8\pi^2\epsilon}\text{Re}\left(\text{Tr} \left[\alpha_{\ell \phi}^\dagger \omega_{\ell \phi D}^{(2)}\right] \right), \\[1em]
		\tilde{c}_{\phi^4 D^4}^{(12)} &\supset \frac{1}{4\pi^2\epsilon}\text{Re}\left(\text{Tr} \left[\alpha_{\ell \phi}^\dagger \omega_{\ell \phi D}^{(2)}\right] \right), \\[1em]
		\tilde{c}_{W\phi^4 D^2}^{(1)} &\supset \frac{g_2}{12\pi^2\epsilon}\text{Re}\left(\text{Tr} \left[2\alpha_{\ell \phi}^\dagger \omega_{\ell \phi D}^{(1)} + \alpha_{\ell \phi}^\dagger \omega_{\ell \phi D}^{(2)}\right] \right), \\[1em]
		\tilde{c}_{W\phi^4 D^2}^{(2)} &\supset \frac{g_2}{8\pi^2\epsilon}\text{Im}\left(\text{Tr} \left[2\alpha_{\ell \phi}^\dagger \omega_{\ell \phi D}^{(1)} + \alpha_{\ell \phi}^\dagger \omega_{\ell \phi D}^{(2)}\right] \right), \\[1em]
		\tilde{c}_{W\phi^4 D^2}^{(3)} &\supset \frac{g_2}{16\pi^2\epsilon}\text{Im}\left(\text{Tr} \left[2\alpha_{\ell \phi}^\dagger \omega_{\ell \phi D}^{(1)} + \alpha_{\ell \phi}^\dagger \omega_{\ell \phi D}^{(2)}\right] \right), \\[1em]
		\tilde{c}_{W\phi^4 D^2}^{(4)} &\supset - \frac{g_2}{16\pi^2\epsilon}\text{Re}\left(\text{Tr} \left[2\alpha_{\ell \phi}^\dagger \omega_{\ell \phi D}^{(1)} + \alpha_{\ell \phi}^\dagger \omega_{\ell \phi D}^{(2)}\right] \right), \\[1em]
		\tilde{c}_{W\phi^4 D^2}^{(5)} &\supset \frac{g_2}{48\pi^2\epsilon}\text{Im}\left(\text{Tr} \left[2\alpha_{\ell \phi}^\dagger \omega_{\ell \phi D}^{(1)} + \alpha_{\ell \phi}^\dagger \omega_{\ell \phi D}^{(2)}\right] \right), \\[1em]
		\tilde{c}_{W\phi^4 D^2}^{(6)} &\supset \frac{g_2}{48\pi^2\epsilon}\text{Re}\left(\text{Tr} \left[2\alpha_{\ell \phi}^\dagger \omega_{\ell \phi D}^{(1)} + \alpha_{\ell \phi}^\dagger \omega_{\ell \phi D}^{(2)}\right] \right), \\[1em]
		\tilde{c}_{W\phi^4 D^2}^{(7)} &\supset \frac{g_2}{48\pi^2\epsilon}\text{Re}\left(\text{Tr} \left[2\alpha_{\ell \phi}^\dagger \omega_{\ell \phi D}^{(1)} + \alpha_{\ell \phi}^\dagger \omega_{\ell \phi D}^{(2)}\right] \right), \\[1em]
		\tilde{c}_{W^2\phi^4}^{(1)} &\supset \frac{g_2^2}{192\pi^2\epsilon}\text{Re}\left(\text{Tr} \left[2\alpha_{\ell \phi}^\dagger \omega_{\ell \phi D}^{(1)} + \alpha_{\ell \phi}^\dagger \omega_{\ell \phi D}^{(2)}\right] \right) - \frac{g_2}{8\pi^2\epsilon}\text{Re}\left(\text{Tr} \left[\alpha_{\ell \phi} \omega_{\ell \phi W}^\dagger \right] \right),\\[1em]
		\tilde{c}_{W^2\phi^4}^{(2)} &\supset \frac{g_2^2}{128\pi^2\epsilon}\text{Im}\left(\text{Tr} \left[2\alpha_{\ell \phi}^\dagger \omega_{\ell \phi D}^{(1)} + \alpha_{\ell \phi}^\dagger \omega_{\ell \phi D}^{(2)}\right]\right)  + \frac{g_2}{8\pi^2\epsilon}\text{Im}\left(\text{Tr} \left[\alpha_{\ell \phi} \omega_{\ell \phi W}^\dagger \right] \right),\\[1em]
		\tilde{c}_{W^2\phi^4}^{(3)} &\supset \frac{g_2^2}{192\pi^2\epsilon}\text{Re}\left(\text{Tr} \left[2\alpha_{\ell \phi}^\dagger \omega_{\ell \phi D}^{(1)} + \alpha_{\ell \phi}^\dagger \omega_{\ell \phi D}^{(2)}\right] \right) + \frac{g_2}{8\pi^2\epsilon}\text{Re}\left(\text{Tr} \left[\alpha_{\ell \phi} \omega_{\ell \phi W}^\dagger \right] \right),\\[1em]
		\tilde{c}_{W^2\phi^4}^{(4)} &\supset \frac{g_2^2}{128\pi^2\epsilon}\text{Im}\left(\text{Tr} \left[2\alpha_{\ell \phi}^\dagger \omega_{\ell \phi D}^{(1)} + \alpha_{\ell \phi}^\dagger \omega_{\ell \phi D}^{(2)}\right] \right) - \frac{g_2}{8\pi^2\epsilon}\text{Im}\left(\text{Tr} \left[\alpha_{\ell \phi} \omega_{\ell \phi W}^\dagger \right] \right),\\[1em]
		\tilde{c}_{WB\phi^4}^{(1)} &\supset \frac{g_1g_2}{96\pi^2\epsilon}\text{Re}\left(\text{Tr} \left[2\alpha_{\ell \phi}^\dagger \omega_{\ell \phi D}^{(1)} + \alpha_{\ell \phi}^\dagger \omega_{\ell \phi D}^{(2)}\right] \right), \\[1em]
		\tilde{c}_{WB\phi^4}^{(2)} &\supset \frac{g_1g_2}{64\pi^2\epsilon}\text{Im}\left(\text{Tr} \left[2\alpha_{\ell \phi}^\dagger \omega_{\ell \phi D}^{(1)} + \alpha_{\ell \phi}^\dagger \omega_{\ell \phi D}^{(2)}\right] \right). 
	\end{align}
	The absent coefficients are not receiving any contribution from LNV operators.

	\subsection{Explicit expression for the RGEs} \label{sec:RGEslist}
	We compute the terms of the anomalous dimension matrix, as discussed at the beginning of this section. The nonzero RGEs for the $d_8$ physical basis operators are:
	\begin{align}\label{eq:rges8}
		\dot{c}_{\phi^8} &= 8 \lambda_\phi  \beta_{\phi D}\, \text{Tr}\left[ \alpha_{\ell \phi} \alpha_{\ell \phi}^\dagger\right] + 32 \lambda_\phi \, \text{Tr}\left[ -\beta_{\phi \ell}^{(1)} \alpha_{\ell \phi} \alpha_{\ell \phi}^\dagger + \beta_{\phi \ell}^{(3)}\alpha_{\ell \phi} \alpha_{\ell \phi}^\dagger \right] \nonumber \\
		&- 16 \lambda_\phi \, \text{Re} \left(\text{Tr} \left[ \alpha_{\ell \phi}^\dagger \omega_{\ell \phi} \right]\right) - \lambda_\phi  g_2^2 \,\text{Re}\left(\text{Tr} \left[2\alpha_{\ell \phi}^\dagger \omega_{\ell \phi D}^{(1)} + \alpha_{\ell \phi}^\dagger \omega_{\ell \phi D}^{(2)}\right] \right) \nonumber \\
		&+  16 \,\text{Tr}\left[\alpha_{\ell \phi} \alpha_{\ell \phi}^\dagger \alpha_{\ell \phi} \alpha_{\ell \phi}^\dagger\right], \\[1em]
		\dot{c}_{\phi^6}^{(1)} &= 4 \beta_{\phi D}\, \text{Tr}\left[\alpha_{\ell \phi} \alpha_{\ell \phi}^\dagger\right] + 16 \, \text{Tr}\left[ 3\beta_{\phi \ell}^{(1)}\alpha_{\ell \phi} \alpha_{\ell \phi}^\dagger - 4 \beta_{\phi \ell}^{(3)} \alpha_{\ell \phi} \alpha_{\ell \phi}^\dagger \right]  \nonumber \\
		&- 16\, \text{Re} \left(\text{Tr} \left[ \alpha_{\ell \phi}^\dagger \omega_{\ell \phi} \right]\right) + \frac{7}{3}g_2^2 \,\text{Re}\left(\text{Tr} \left[2\alpha_{\ell \phi}^\dagger \omega_{\ell \phi D}^{(1)}+ \alpha_{\ell \phi}^\dagger \omega_{\ell \phi D}^{(2)}\right] \right) \nonumber\\
		&- 32\lambda_\phi  \text{Re}\left(\text{Tr} \left[\alpha_{\ell \phi}^\dagger \omega_{\ell \phi D}^{(2)}\right] \right)- 4 \,\text{Im}\left(\text{Tr} \left[\alpha_{\ell \phi} y^e \omega_{\ell \phi D e}^* \right] \right) \nonumber \\
		&- 2\,\text{Re}\left(\text{Tr} \left[\alpha_{\ell \phi}^\dagger \left(y^e\right)^* \left(y^e\right)\left(\omega_{\ell \phi D}^{(1)} + \left(\omega_{\ell \phi D}^{(1)}\right)^T \right)  \right] \right) \nonumber\\
		& - 2\, \text{Re}\left(\text{Tr} \left[\alpha_{\ell \phi}^\dagger \left(y^e\right)^* \left(y^e\right) \omega_{\ell \phi D}^{(2)}\right] \right), \\[1em]
		\dot{c}_{\phi^6}^{(2)} &=8 \beta_{\phi D}\, \text{Tr}\left[\alpha_{\ell \phi}\alpha_{\ell \phi}^\dagger\right] +16 \, \text{Tr}\left[ 2\beta_{\phi \ell}^{(1)}\alpha_{\ell \phi} \alpha_{\ell \phi}^\dagger - \beta_{\phi \ell}^{(3)}\alpha_{\ell \phi} \alpha_{\ell \phi}^\dagger \right]  \nonumber \\
		&- 8\, \text{Re} \left(\text{Tr} \left[\alpha_{\ell \phi}^\dagger \omega_{\ell \phi}\right]\right) + \frac{1}{12}g_2^2 \,\text{Re}\left(\text{Tr} \left[2\alpha_{\ell \phi}^\dagger \omega_{\ell \phi D}^{(1)} + \alpha_{\ell \phi}^\dagger \alpha_{\ell \phi D}^{(2)} \right] \right)\nonumber\\
		& - 16\lambda_\phi  \,\text{Re}\left(\text{Tr} \left[\alpha_{\ell \phi}^\dagger \alpha_{\ell \phi D}^{(2)} \right] \right)+ 4 \,\text{Im}\left(\text{Tr} \left[\alpha_{\ell \phi} \left(y^e\right) \omega_{\ell \phi D e}^* \right] \right) \nonumber \\
		&+ 2\,\text{Re}\left(\text{Tr} \left[\alpha_{\ell \phi} \left(y^e\right)^* \left(y^e\right) \left(\omega_{\ell \phi D}^{(1)} + \left(\omega_{\ell \phi D}^{(1)}\right)^T \right)  \right] \right) \nonumber\\
		& + 2\, \text{Re}\left(\text{Tr} \left[\alpha_{\ell \phi}^\dagger \left(y^e\right)^* \left(y^e\right) \omega_{\ell \phi D}^{(2)} \right] \right), \label{eq:rgesphi62} \\[1em]
		\dot{c}_{\phi^4}^{(2)} &= -8\,\text{Re}\left(\text{Tr} \left[\alpha_{\ell \phi}^\dagger \omega_{\ell \phi D}^{(2)}\right] \right),\label{eq:rgesphi4d4}\\[1em]
		\dot{c}_{W\phi^4 D^2}^{(1)} &= -4g_2\,\text{Re}\left(\text{Tr} \left[2\alpha_{\ell \phi}^\dagger \omega_{\ell \phi D}^{(1)} + \alpha_{\ell \phi}^\dagger \omega_{\ell \phi D}^{(2)}\right] \right), \\[1em]
		\dot{c}_{W\phi^4 D^2}^{(2)} &= -4g_2\,\text{Im}\left(\text{Tr} \left[2\alpha_{\ell \phi}^\dagger \omega_{\ell \phi D}^{(1)} + \alpha_{\ell \phi}^\dagger \omega_{\ell \phi D}^{(2)}\right] \right), \\[1em]
		\dot{c}_{W\phi^4 D^2}^{(3)} &= -2g_2\,\text{Im}\left(\text{Tr} \left[2\alpha_{\ell \phi}^\dagger \omega_{\ell \phi D}^{(1)} + \alpha_{\ell \phi}^\dagger \omega_{\ell \phi D}^{(2)}\right] \right), \\[1em]
		\dot{c}_{W\phi^4 D^2}^{(4)} &= 2g_2\,\text{Re}\left(\text{Tr} \left[2\alpha_{\ell \phi}^\dagger \omega_{\ell \phi D}^{(1)} + \alpha_{\ell \phi}^\dagger \omega_{\ell \phi D}^{(2)}\right] \right), \\[1em]
		\dot{c}_{W^2\phi^4}^{(1)} &= -\frac{1}{4}g_2^2\,\text{Re}\left(\text{Tr} \left[2\alpha_{\ell \phi}^\dagger \omega_{\ell \phi D}^{(1)} + \alpha_{\ell \phi}^\dagger \omega_{\ell \phi D}^{(2)}\right] \right) + 4g_2\,\text{Re}\left(\text{Tr} \left[\alpha_{\ell \phi} \omega_{\ell \phi W}^\dagger\right] \right), \\[1em]
		\dot{c}_{W^2\phi^4}^{(2)} &= -\frac{1}{4}g_2^2\,\text{Im}\left(\text{Tr} \left[2\alpha_{\ell \phi}^\dagger \omega_{\ell \phi D}^{(1)} + \alpha_{\ell \phi}^\dagger \omega_{\ell \phi D}^{(2)}\right] \right) - 4g_2\,\text{Im}\left(\text{Tr} \left[\alpha_{\ell \phi}\omega_{\ell \phi W}^\dagger\right] \right), \\[1em]
		\dot{c}_{W^2\phi^4}^{(3)} &= -\frac{1}{4}g_2^2\,\text{Re}\left(\text{Tr} \left[2\alpha_{\ell \phi}^\dagger \omega_{\ell \phi D}^{(1)} + \alpha_{\ell \phi}^\dagger \omega_{\ell \phi D}^{(2)} \right] \right) - 4g_2\,\text{Re}\left(\text{Tr} \left[\alpha_{\ell \phi} \omega_{\ell \phi W}^\dagger\right] \right), \\[1em]
		\dot{c}_{W^2\phi^4}^{(4)} &= -\frac{1}{4}g_2^2\,\text{Im}\left(\text{Tr} \left[2\alpha_{\ell \phi}^\dagger \omega_{\ell \phi D}^{(1)} +\alpha_{\ell \phi}^\dagger \omega_{\ell \phi D}^{(2)}\right] \right) + 4g_2\,\text{Im}\left(\text{Tr} \left[\alpha_{\ell \phi}\omega_{\ell \phi W}^\dagger\right] \right), \\
		\dot{c}_{WB\phi^4}^{(1)} &= -g_1g_2\,\text{Re}\left(\text{Tr} \left[\alpha_{\ell \phi}^\dagger \omega_{\ell \phi D}^{(1)}\right] \right) -\frac{1}{2}g_1g_2 \,\text{Re}\left(\text{Tr} \left[\alpha_{\ell \phi}^\dagger \omega_{\ell \phi D}^{(2)}\right] \right), \\
		\dot{c}_{WB\phi^4}^{(2)} &= -g_1g_2\,\text{Im}\left(\text{Tr} \left[\alpha_{\ell \phi}^\dagger \omega_{\ell \phi D}^{(1)}\right] \right) -\frac{1}{2}g_1g_2 \,\text{Im}\left(\text{Tr} \left[\alpha_{\ell \phi}^\dagger \omega_{\ell \phi D}^{(2)}\right] \right).
	\end{align}%
	As discussed in Section~\ref{sec:belowd8}, there is no renormalisation at order \order{4} coming directly from the 1PI diagrams with LNV vertices for any coefficient of dimension less than eight. However, the redundancies of the operators in the Green's basis do contribute to these equations. All of these yield a non-vanishing RGE for some coefficients. Again, we only show the nonzero RGEs:
	\begin{align}\label{eq:rges6}
		\dot{\lambda}_\phi &= -8 \mu^4_\phi \,\text{Re}\left(\text{Tr} \left[\alpha_{\ell \phi}^\dagger \omega_{\ell \phi D}^{(2)}\right] \right),\\
		\dot{c}_{\phi} &= -4 \mu^2_\phi \beta_{\phi D}\, \text{Tr}\left[\alpha_{\ell \phi} \alpha_{\ell \phi}^\dagger\right] + 16 \mu^2_\phi \, \text{Tr}\left[ \beta_{\phi \ell}^{(1)} \alpha_{\ell \phi} \alpha_{\ell \phi}^\dagger \right] \nonumber \\
		&- 16 \mu^2_\phi \, \text{Tr}\left[\beta_{\phi \ell}^{(3)} \alpha_{\ell \phi} \alpha_{\ell \phi}^\dagger \right] + 8 \mu^2_\phi\, \text{Re} \left(\text{Tr} \left[ \alpha_{\ell \phi}^\dagger \omega_{\ell \phi} \right]\right) \nonumber \\
		& + \mu^2_\phi  g_2^2 \,\text{Re}\left(\text{Tr} \left[\alpha_{\ell \phi}^\dagger \omega_{\ell \phi D}^{(1)} \right]\right) + \frac{1}{2}\mu^2_\phi g_2^2 \,\text{Re}\left(\text{Tr} \left[\alpha_{\ell \phi}^\dagger \omega_{\ell \phi D}^{(2)} \right] \right) \nonumber \\
		&+ 16 \lambda_\phi  \mu^2_\phi \,\text{Re}\left(\text{Tr} \left[\alpha_{\ell \phi}^\dagger \omega_{\ell \phi D}^{(2)} \right] \right), \\
		\dot{c}_{\phi \square} &= 16 \mu^2_\phi \,\text{Re}\left(\text{Tr} \left[\alpha_{\ell \phi}^\dagger \omega_{\ell \phi D}^{(2)}\right] \right), \\
		\dot{c}_{\phi D} &= 16 \mu^2_\phi \,\text{Re}\left(\text{Tr} \left[\alpha_{\ell \phi}^\dagger \omega_{\ell \phi D}^{(2)}\right] \right).\label{eq:rgecphid}
	\end{align}
	
	The previous results for the RGEs are expressed in Table~\ref{tab:anomalousdim}. There, it is easy to check the numerical contribution for each operator coming from all the possible sources detailed previously. In cases where there is more than one term for the same contribution, we show only the greatest. Naive loop suppression $\gamma\sim 1$ competes with a high numerical coefficient for some of the terms, especially those of class $\phi^6 D^2$, which hold the greatest terms of all the renormalised WC.
	
	As discussed in section~\ref{sec:sortingcontribs}, we see many vanishing contributions in Table~\ref{tab:anomalousdim}. All of the entries are well understood in terms of the logic stated before: the order of the operator expansion and loop expansion altogether with the restriction to bosonic operators leaves little scope for non-vanishing anomalous dimension matrix elements. Nevertheless, there are two cases worth remarking here, after the computations are done.
	
	First of all, the nonrenormalised WC of $\mathcal{O}_{\phi^4}^{(2)}$ by $\alpha_{\ell \phi}\omega^{(1)}_{\ell\phi D}$ is the result of an accidental zero. The contributions from class $\psi^2\phi^2 D^2$ to $\phi^4$ are not forbidden by any theoretical argument and in fact they are expected to be non-zero like the contribution from $\alpha_{\ell \phi}\omega^{(2)}_{\ell\phi D}$. In this case, the reason lies in the operator structure, $\mathcal{O}^{(1)}_{\ell\phi D} = \epsilon_{ij}\epsilon_{mn}\ell^iC(D^\mu \ell^j)\phi^m (D_\mu\phi^n) $, which under the contraction with the Weinberg operator to make a 4-Higgs process is anti-symmetric under swapping of the internal loop propagators $ (\ell^i \leftrightarrow \ell^j) $ and thus the amplitude vanishes. This is an artefact of both the operator and the process, as in contrast, the mixing contribution from $\mathcal{O}^{(2)}_{\ell\phi D}$ is non-vanishing.
	
	All other non-trivial zeros come from the class $X \phi^2 D^2$. However, in this case, the diagrams \ref{fig:d5_psi2phi2X} vanish regardless of the internal structure of the operator. This can be demonstrated through unitary cuts but also considering that there is only one independent operator for each subclass $B \phi^2 D^2$ and $W \phi^2 D^2$ which means the symmetry argument discussed in section \ref{sec:sortd5d7} will hold after any possible redefinition of operators $\mathcal{O}_{\ell\phi B}$ and $\mathcal{O}_{\ell\phi W}$.
	
	\begin{table}[htb!]
		\begin{center}
			\resizebox{\textwidth}{!}{
				\begin{tabular}{cccccccccc}
					& $\left(\alpha_{\ell \phi}\right)^4$ & $\left(\alpha_{\ell \phi}\right)^2 \beta_{\phi D}$ & $\left(\alpha_{\ell \phi}\right)^2 \beta_{\phi \ell}^{(1)}$ & $\left(\alpha_{\ell \phi}\right)^2 \beta_{\phi \ell}^{(3)}$ & $\alpha_{\ell \phi} \omega_{\ell \phi} $ & $\alpha_{\ell \phi} \omega_{\ell \phi D}^{(1)} $ & $\alpha_{\ell \phi} \omega_{\ell \phi D}^{(2)} $ & $\alpha_{\ell \phi} \omega_{\ell \phi D e} $ & $\alpha_{\ell \phi} \omega_{\ell \phi W} $ \cr
					\hline
					$\gamma_{\phi^8}$ & 16 & $8 \lambda_\phi $ & $32 \lambda_\phi $ & $32 \lambda_\phi $ & $16 \lambda_\phi $ & $2  \lambda_\phi  g_2^2$ & $\lambda_\phi  g_2^2$ & 0 & 0 \\
					$\gamma_{\phi^6}^{(1)}$ & 0 & 4 & 48 & 64 & 16 & $\frac{14}{3}g_2^2$ & $32 \lambda_\phi $ & $4 y^e$ & 0 \cr 
					$\gamma_{\phi^6}^{(2)}$ & 0 & 8 & 32 & 16 & 8 & $\frac{1}{6}g_2^2$ & $16 \lambda_\phi $ & $4 y^e$ & 0 \cr 
					$\gamma_{\phi^4}^{(2)}$ & 0 & 0 & 0 & 0 & 0 & $\slashed{0}$ & $8$ & 0 & 0 \cr
					$\gamma_{W\phi^4 D^2}^{(1)}$ & 0 & 0 & 0 & 0 & 0 & $8 g_2$ & $4 g_2$ & 0 & $\slashed{0}$ \cr
					$\gamma_{W\phi^4 D^2}^{(2)}$ & 0 & 0 & 0 & 0 & 0 & $8 g_2$ & $4 g_2$ & 0 & $\slashed{0}$ \cr
					$\gamma_{W\phi^4 D^2}^{(3)}$ & 0 & 0 & 0 & 0 & 0 & $4 g_2$ & $2 g_2$ & 0 & $\slashed{0}$ \cr
					$\gamma_{W\phi^4 D^2}^{(4)}$ & 0 & 0 & 0 & 0 & 0 & $4 g_2$ & $2 g_2$ & 0 & $\slashed{0}$ \cr
					$\gamma_{W^2 \phi^4}^{(1)}$ & 0 & 0 & 0 & 0 & 0 & $ \frac{1}{2} g_2^2$ & $ \frac{1}{4} g_2^2$ & 0 & $4 g_2$ \cr
					$\gamma_{W^2 \phi^4}^{(2)}$ & 0 & 0 & 0 & 0 & 0 & $ \frac{1}{2} g_2^2$ & $ \frac{1}{4} g_2^2$ & 0 & $4 g_2$ \cr
					$\gamma_{W^2 \phi^4}^{(3)}$ & 0 & 0 & 0 & 0 & 0 & $ \frac{1}{2} g_2^2$ & $ \frac{1}{4} g_2^2$ & 0 & $4 g_2$ \cr
					$\gamma_{W^2 \phi^4}^{(4)}$ & 0 & 0 & 0 & 0 & 0 & $ \frac{1}{2} g_2^2$ & $ \frac{1}{4} g_2^2$ & 0 & $4 g_2$ \cr
					$\gamma_{WB \phi^4}^{(1)}$ & 0 & 0 & 0 & 0 & 0 & $g_1 g_2$ & $\frac{1}{2}g_1g_2$ & 0 & $\slashed{0}$ \cr
					$\gamma_{WB \phi^4}^{(2)}$ & 0 & 0 & 0 & 0 & 0 & $g_1 g_2$ & $\frac{1}{2}g_1g_2$ & 0 & $\slashed{0}$ \cr \\[-0.5cm]
					\hline 
					$\gamma_{\phi}$ & 0 & $4\mu^2_\phi$ & $16\mu^2_\phi$ & $16\mu^2_\phi$ & $8\mu^2_\phi$ & $\mu^2_\phi g_2^2$ & $16\mu^2_\phi \lambda_\phi $ & 0 & 0 \cr
					$\gamma_{\phi \square}$ & 0 & 0 & 0 & 0 & 0 & 0 & $16\mu^2_\phi$ & 0 & 0 \cr
					$\gamma_{\phi D}$ & 0 & 0 & 0 & 0 & 0 & 0 & $16\mu^2_\phi$ & 0 & 0 \cr
					$\gamma_{\lambda_\phi }$ & 0 & 0 & 0 & 0 & 0 & 0 & $8\mu^4_\phi$ & 0 & 0 \cr
			\end{tabular}}
		\end{center}
		\caption{\label{tab:anomalousdim}Anomalous dimension matrix. The columns represent the greatest terms from each contribution in absolute value that renormalise the coefficients in the rows. See equations \ref{eq:rges8} -- \ref{eq:rgecphid} for complete RGEs. $0$ represents a trivially vanishing contribution due to the absence of Feynman diagrams. $\slashed{0}$ represents non-trivially vanishing contribution. See the discussion in Section \ref{sec:sortingcontribs}.}
	\end{table}
	
	\subsection{Detailed example} \label{sec:exampleRGE}
	Let us work out a detailed example of this calculation for clarity: the contribution of all the LNV operators to the RGE of $c_{\phi^4}^{(2)}$. First, we implement the Lagrangian Eq.~\ref{eq:LagUV} in a \texttt{FeynRules} model. With this, we generate the topologies and processes in \texttt{FeynArts}, which -paired with \texttt{FormCalc}- computes the amplitudes for the different processes. In particular, for $c_{\phi^4}^{(2)}$ we get information from the process $\phi^0 \rightarrow \phi^0 \phi^+ \phi^- $. Of course, this is not the only process that could yield the divergence for this particular coefficient. Generally, different processes give the divergences for the same set of coefficients, so it is a matter of finding a suitable process in each case. In our example, we have
	
	\begin{align}
		-\text{i}\mathcal{A}_{IR}&=2c_{\phi^4}^{(1)}\left(-\kappa_{1213}+\kappa_{1322}+\kappa_{1323}\right)+2c_{\phi^4}^{(2)}\left(-\kappa_{1123}+\kappa_{1223}+\kappa_{1323}\right)\nonumber\\
		&+2c_{\phi^4}^{(3)}\left(-\kappa_{1213}+\kappa_{1223}+\kappa_{1233}\right)+2c_{\phi^4}^{(4)}\left(\kappa_{1212}+\kappa_{1213}-\kappa_{1223}-\kappa_{1233}\right)\nonumber\\
		&+c_{\phi^4}^{(4)}\left(-\kappa_{1112}+\kappa_{1113}-\kappa_{1123}-\kappa_{1133}-\kappa_{1222}+\kappa_{1322}-\kappa_{2223}-\kappa_{2233}\right)\nonumber\\
		&+c_{\phi^4}^{(6)}\left(\kappa_{1112}-\kappa_{1113}-\kappa_{1122}-\kappa_{1123}+\kappa_{1233}-\kappa_{1333}-\kappa_{2223}-\kappa_{2333}\right)\nonumber\\
		&+2c_{\phi^4}^{(6)}\left(\kappa_{1313}+\kappa_{1213}-\kappa_{1322}-\kappa_{1323}\right)+4c_{\phi^4}^{(8)}\left(-\kappa_{1112}-\kappa_{1113}+\kappa_{1123}\right)\nonumber\\
		&+2c_{\phi^4}^{(8)}\left(\kappa_{1111}+\kappa_{1122}+\kappa_{1133}+\kappa_{2233}\right)+c_{\phi^4}^{(10)}\left(\kappa_{3333}+\kappa_{1122}+\kappa_{1133}+\kappa_{2233}\right)\nonumber\\
		&+2c_{\phi^4}^{(10)}\left(-\kappa_{1233}-\kappa_{1333}+\kappa_{2333}\right)+c_{\phi^4}^{(11)}\left(\kappa_{2222}+\kappa_{1122}+\kappa_{1133}+\kappa_{2333}\right)\nonumber\\
		&+2c_{\phi^4}^{(11)}\left(-\kappa_{1222}-\kappa_{1322}+\kappa_{2223}\right)+2c_{\phi^4}^{(12)}\left(\kappa_{2323}+\kappa_{1123}-\kappa_{1223}-\kappa_{1323}\right)\nonumber\\&+c_{\phi^4}^{(12)}\left(-\kappa_{1122}-\kappa_{1133}+\kappa_{1222}+\kappa_{1233}+\kappa_{1322}+\kappa_{1333}+\kappa_{2223}+\kappa_{2333}\right),
	\end{align}
	where $\kappa_{ijkl}=p_i\cdot p_j p_k\cdot p_l$ are the kinematic invariants, and we do not include CP-odd terms since we already know they do not contribute to the CP-even $c_{\phi^4}^{(2)}$. This amplitude has been built at tree level, including redundant operators in the $d_8$ Green's basis so that only 1PI diagrams are considered here, as explained in Section~\ref{sec:theory}. Consequently, with this method, the amplitude is off-shell. Now for the 1-loop amplitude, we have:
	\begin{equation}
		-\text{i}\mathcal{A}_{UV} =\frac{1}{8\pi^2\epsilon}\text{Re}\left[\alpha_{\ell \phi}^\dagger \omega_{\ell \phi D}^{(2)}\right]\left(\kappa_{2222}+\kappa_{3333}+2\kappa_{2233}+4\kappa_{2323}+4\kappa_{2333}\right),
	\end{equation}
	
	which indicates that the only contribution for this process comes from the 4-Higgs diagram displayed in Figure~\ref{fig:d5_psi2phi2D2}.
	
	Equating the two amplitudes matches the Wilson coefficients effectively to their divergences. The nonzero solutions for our example are:
	\begin{equation}
		c_{\phi^4}^{(2)}=2c_{\phi^4}^{(10)}=2c_{\phi^4}^{(11)}=c_{\phi^4}^{(12)}=\frac{1}{4\pi^2\epsilon}\text{Re}\left[\alpha_{\ell \phi}^\dagger \omega_{\ell \phi D}^{(2)}\right]\,.
	\end{equation}
	
	We have obtained the divergences of $c_{\phi^4}^{(2)}$, but following the off-shell method, we also need the divergences for the redundant coefficients that appear in the redefinition of $c_{\phi^4}^{(2)}$.
	According to Ref.~\cite{Chala:2021cgt}, the shift for this coefficient is:
	\begin{equation} \label{eq:wcshiftphi42}
		c_{\phi^4}^{(2)}\rightarrow c_{\phi^4}^{(2)}+g_1c_{B\phi^2D^4}^{(3)}+g_2c_{W\phi^2D^4}^{(3)}-g_1^2c_{B^2D^4}-g_2^2c_{W^2D^4}.
	\end{equation}
	Applying all the divergences to Eq.~\ref{eq:wcshiftphi42}, we get the physical 1-loop coefficient:
	\begin{equation} \label{eq:physicalphi42}
		\tilde{c}_{\phi^4}^{(2)}=\frac{1}{4\pi^2\epsilon}\text{Re}\left[\alpha_{\ell \phi}^\dagger \omega_{\ell \phi D}^{(2)}\right]\,.
	\end{equation}
	Now, for equation Eq.~\ref{eq:rgedef}, we need the tree-level anomalous dimension $n_j$ of operators $\alpha_{\ell \phi}$ and $\omega_{\ell \phi D}^{(2)}$. One can trivially check that 
	\begin{equation}
		\text{Mass dim.}\left[\mu^{-(n_{\ell \phi}^{(5)}) \epsilon}\frac{\alpha_{\ell \phi}}{\Lambda}\mathcal{O}_{\ell \phi}^{(5)}\right]\mapsto 4-(n_{\ell \phi}^{(5)})\epsilon =4-2\epsilon \Rightarrow n_{\ell \phi}^{(5)}=2 .
	\end{equation}
	And through a similar procedure $n_{\ell \phi D}^{(2)}=3$. 
	Finally, all that remains is to use the definition of the RGEs Eq.~\ref{eq:rgedef}, which is straightforward since we have all the ingredients needed. Note that $x_j$ in the formula comprehends \textit{all} the coefficients, running or not, which includes all the WCs, but also the SM couplings and masses $\mu^2_\phi$, $\lambda_\phi $, $g_1$, $g_2$, and $g_3$. In this case, the derivative is trivial, and so the RGE reads:
	\begin{equation}
		\dot{c}_{\phi^4}^{(2)}=-8\,\text{Re}\left[\alpha_{\ell \phi}^\dagger \omega_{\ell \phi D}^{(2)} \right]\,.
	\end{equation}
	There are other contributions to Eq.~\ref{eq:physicalphi42} at order \order{4}, which are from lepton number conserving operators. The rest of the terms, those coming from other $d_8$ operators or pairs of $d_6$, are explicitly shown in Refs.~\cite{Chala:2021pll,DasBakshi:2022mwk}.

	\section{Discussions and outlook}\label{sec:Disc}
	Here, we discuss some implications on BSM parameter space based on these RGEs.
	
	\subsection{Positivity bounds in seesaw models}\label{subsec:pos}
	Positivity bounds are restrictions on the S-matrix elements derived using the standard axioms -- analyticity, unitarity, and crossing symmetry on the underlying theory. In the context of SMEFT, these bounds constrain the space of Wilson coefficients of $d_8$ operators. The positivity bounds for $ 2 \rightarrow 2 $ scattering amplitude $ \mathcal{A}(s,t)$ in the forward  scattering limit\,$ (t=0) $ are deduced using \cite{Adams:2006sv},
	\begin{align}\label{eq:positivitydef}
		\left.\frac{d^2}{ds^2} \mathcal{A}(s,0)\right|_{s=0} \geq 0.
	\end{align}
	Assuming that $ A(s,0) $ is analytic around the origin and is expanded around $ s=0 $,
	\begin{align}
		A(s,0) = a_0 + a_1 s + a_2 s^2 + \ldots \, ,
	\end{align}
	which implies that $ a_2 \geq 0 $ following Eq.~\ref{eq:positivitydef}. $ a_2  $ is defined by combinations of $d_8$ operators (or pairs of $d_6$ operators), and the positivity bounds restrict these combinations. For example, the process $\phi \phi \rightarrow \phi \phi$ yields positivity bounds affecting the Wilson coefficients of $\phi^4D^4$ operators\cite{Remmen:2019cyz},
	
	\begin{align}
		\phi_1 \phi_2 \rightarrow \phi_1 \phi_2 &: c^{(2)}_{\phi^4}  \geq 0 \ , \nonumber \\
		\phi_1 \phi_3 \rightarrow \phi_1 \phi_3 &:  c^{(1)}_{\phi^4}+c^{(2)}_{\phi^4}  \geq 0 \ , \\
		\phi_1 \phi_1 \rightarrow \phi_1 \phi_1 &: c^{(1)}_{\phi^4}+c^{(2)}_{\phi^4}+c^{(3)}_{\phi^4}  \geq 0 \ . \nonumber
	\end{align}
	
	The amplitude around $ s=0 $ can be irregular in certain cases, such as contributions from loops carrying massless propagators, where one encounters branch cuts extending to the origin. Generally, one arrives at Eq.~\ref{eq:positivitydef} by lending a small mass `$ m $' to the massless state to regularise the singularity at the origin and resetting it to zero at a later stage. Nonetheless, one needs to scrutinise this procedure when loops with massless propagators are present \cite{Chala:2021wpj}.
	
	For $ \phi^4 D^4 $ operators, the running effects of lower dimensional operators ($\phi^4 $ and $ \phi^4 D^2 $) dominate the corresponding amplitude in the limit $ m \rightarrow 0 $ \cite{Chala:2021wpj}. The positivity bounds on $  \phi^4 D^4  $ class are respected by the running of effective interactions, which implies the $ d_8 $ RGEs restrict arbitrary values of the WCs provided that $\phi^4 $ and $ \phi^4 D^2 $ are absent at the tree-level. For instance, we can consider the bound $ c_{\phi^4}^{(2)}(\Lambda) \geq 0$ and 
	Eq.~\ref{eq:rgesphi4d4}, then, Eq.~\ref{eq:rgedefinition} evaluates to
	\begin{align}\label{eq:c5ctbound}
		16 \pi^2 c_{\phi^4}^{(2)}(\mu) =  8\,\text{Re}\left(\text{Tr} \left[\alpha_{\ell \phi}^\dagger \omega_{\ell \phi D}^{(2)}\right]  \right) \text{ln}\frac{\Lambda}{\mu} & \geq 0 \nonumber \\
		\Rightarrow \ \   \text{Re}\left(\text{Tr} \left[\alpha_{\ell \phi}^\dagger \omega_{\ell \phi D}^{(2)}\right]  \right) & \geq 0\,,
	\end{align}
	for $\mu < \Lambda$ in the limit of scale-invariance of the LNVs. This inequality imposes significant restrictions on new physics parameter space. Specifically, if we assume the WCs are nonzero for one flavour index and are real-valued, then $ \alpha_{\ell \phi,mm} $ and $ \omega_{\ell \phi D,mm}^{(2)} $ must have the same signs in all possible UV completions of SMEFT that do not generate $ \phi^4 D^4 $ WCs at tree level. For three flavour indices, the restriction is
	\begin{align}\label{eq:c5ctboundallgeneration}
		\sum_{mn}^3  \alpha_{\ell \phi,mn}  \omega_{\ell \phi D,mn}^{(2)} \geq 0.
	\end{align}
	We emphasise that this restriction is deduced using purely IR information $, i.e. $ the RGE calculation in the low energy theory (SMEFT in our case), and the positivity bounds impose direct constraints on the UV space.
	
	Let us inspect this inequality in BSMs where these LNVs are generated at tree-level matching, but $ \phi^4D^4 $ class of $ d_8 $ operators are not. A suitable choice for such BSMs are Type-I and III seesaw models \cite{Mohapatra:1979ia,Malinsky:2005bi,Du:2022vso}.
	
	We add two heavy multiplets to the SM particle content $N^i : (1,1,0)_{\frac{1}{2}}$  and $\Sigma^i : (1,3,0)_{\frac{1}{2}} $ with SM gauge and spin quantum numbers shown as: $ (SU(3)_C , SU(2)_L , U(1)_Y)_{S} $. The BSM Lagrangian is defined as
	\begin{align}\label{eq:BSMLag}
		\mathcal{L}  = & \ \mathcal{L}_{\text{SM}} + \frac{1}{2} \bar{N} (i \slashed{\partial} - M_N )  N
		%
		%
		+ \frac{1}{2} \text{Tr}[\bar{\Sigma} (i \slashed{D} - M_\Sigma)  \Sigma] \nonumber\\
		%
		%
		&- \frac{1}{2} \left\lbrace Y^{N \dagger}_{ij}  \, \bar{N}^i \, \widetilde{\phi}^\dagger \, \ell^j 
		+ (Y^{N})^T_{ij}  \, \bar{N}^i \, \widetilde{\phi}^T \, \ell^{Cj}
		%
		%
		%
		+ Y^\Sigma_{ij}  \, \widetilde{\phi}^\dagger \bar{\Sigma}^i \, \ell^j
		+ Y^{\Sigma \ast}_{ij}  \, \phi^\dagger \bar{\Sigma}^i \, i \sigma_2 \ell^{Cj}
		\text{+ h.c.} \right\rbrace,
	\end{align}
	where $\Sigma$  is a $SU(2)$ triplet expressed in 2-dimensional matrix form.
	We integrate out the heavy fields at tree-level onto SMEFT up to order~\order{3}, leading to the $d_7$ operator $ \mathcal{O}'_{7ij} $ and its WC $\mathcal{C}_{7ij}'$,
	
	\begin{align}
		\mathcal{O}_{7ij}'&= (\bar{\ell}^C_{i} \tilde{\phi}^\ast) \slashed{D}^2(\tilde{\phi}^\dagger \ell_{j}) \ , \
		\mathcal{C}_{7ij}'= -\frac{\left[(Y^N)^\ast (Y^N)^\dagger\right]_{ij}}{2 M_N^{3}} -\frac{\left[(Y^{\Sigma})^{T} (Y^\Sigma)\right]_{ij}}{2 M_\Sigma^{3}} \, , 
	\end{align}
	
	$ \mathcal{O}'_{7ij} $ is related to operators in the basis we have considered (Table~\ref{tab:op67}). Using the SM equation of motions and properties of Dirac-gamma matrices, one can reduce it to operators of classes $ \psi^2\phi^2 $, $ \psi^2\phi^4 $, $ \psi^4\phi$, $ \psi^2\phi^3 D $, and $ \psi^2 \phi^2 X $. Note that of all $ d_7 $ operators only $ \mathcal{O}_{\ell\phi D}^{(2)} $ renormalises the $ \phi^4 D^4 $ class operators as can be seen in the only nonzero RGE of that class Eq.~\ref{eq:rgesphi4d4}. For these models, $\mathcal{O}_{\ell\phi D}^{(2)}$ is not generated from tree-level matching, and so, $ \omega_{\ell\phi D}^{(2)}=0 $. Therefore, we arrive at the important conclusion that the restrictions derived from positivity bounds in Eqs.~\ref{eq:c5ctbound} and \ref{eq:c5ctboundallgeneration} are respected in these seesaw models.
	
	\subsection{Constraints from T-parameter}
\begin{figure}[h]
		\begin{center}
			\includegraphics[scale=0.5]{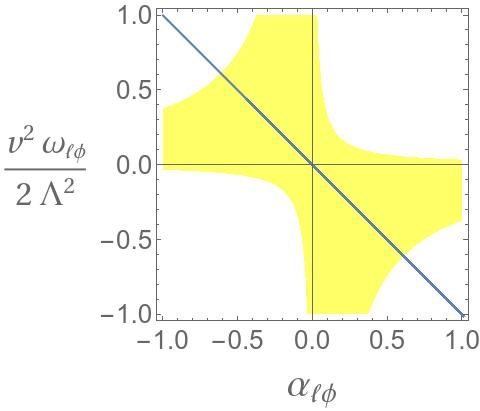}
		\end{center}\vspace{-.5cm}
		\caption{The restrictions imposed on Weinberg-like operators, {\footnotesize $\mathcal{O}_{\ell \phi}^{(5)}$} at $d_5$ and {\footnotesize $\mathcal{O}_{\ell \phi}^{(7)}$} at $d_7$, translated from bounds on the T-parameter for one non-vanishing flavour direction. Neutrino mass bounds restrict these two Wilson coefficients to lie on the shown blue contour, whereas the T-parameter bound restricts the Wilson coefficients within the yellow region. The plot is created for $\Lambda= 1 $ TeV, and $\mu = 246 $ GeV, with inputs $ T = 0.10 \pm 0.12 $ \cite{deblas:2016nqo}, and $ M_N < 0.081 $ eV \cite{Loureiro:2018pdz}.}\label{fig:Plot2}
	\end{figure}

	Here, using the RGE relations presented in subsection~\ref{sec:RGEslist}, we discuss the restrictions imposed by the T-parameter on the LNVs via the $d_8$ class $ \phi^6 D^2 $.
	The Wilson coefficients $ \beta_{\phi D} $ ($d_6$) and $ c^{(2)}_{\phi^6} $ ($d_8$) contribute to the T-parameter \cite{deblas:2016nqo}:
	\begin{align}\label{eq:Tparambounds}
		T = -\frac{1}{2 \alpha } \frac{v^2}{\Lambda^2}\left(\beta_{\phi D} + c^{(2)}_{\phi^6} \frac{v^2}{\Lambda^2} \right),
	\end{align}
	where $\alpha \sim \frac{1}{137}$ is the fine structure constant. The LNVs renormalise $ \mathcal{O}^{(2)}_{\phi^6} $, and therefore, the T-parameter restricts these WCs from arbitrary values. This becomes important where $ \phi^4 D^2 $ and $ \phi^6 D^2 $ classes do not appear at tree-level. In such case, from the RGE of $ c^{(2)}_{\phi^6} $ (Eq.~\ref{eq:rgesphi62}) and from Eq.~\ref{eq:Tparambounds}, we get (in the limit of scale-invariant WCs),
	\begin{align}
		T = -\frac{1}{4 \pi^2 \alpha } \frac{v^4}{\Lambda^4} \ln \left[\frac{\Lambda}{\mu}\right] \sum_{mn}^3 \alpha_{\ell \phi,mn} \omega_{\ell \phi,mn},
	\end{align}
	assuming that only the WCs of Weinberg-like operators ($ \alpha_{\ell \phi} $ and $ \omega_{\ell \phi} $) are non-vanishing and real valued. This puts restrictions on the 
	$ \alpha_{\ell \phi} - \omega_{\ell \phi} $ plane, which can be compared to the bounds from neutrino masses \cite{Loureiro:2018pdz},
	\begin{align}
		\left(M_{N}\right)_{mn} = -\frac{v^2}{\Lambda} \left(\alpha_{\ell \phi,mn} + \frac{v^2}{2 \Lambda^2} \omega_{\ell \phi,mn} \right).
	\end{align}
	It is straightforward to visualise the restrictions on $ \alpha_{\ell \phi} $ vs $ \omega_{\ell \phi} $ plane taking both the T-parameter and neutrino mass bounds in the case of single non-vanishing flavor direction as shown in Fig.~\ref{fig:Plot2}. Note that, in general, neutrino masses do not restrain $ \alpha_{\ell \phi} $ and $ \omega_{\ell \phi} $ to possess large values. This blind direction is lifted by the T-parameter, as it imposes an upper bound to these WCs on the account of the RGEs of $d_8$ operators, which is shown in Fig.~\ref{fig:Plot2}. %
	\section{Conclusions}\label{sec:Conclusion}
	
		\begin{table}[htb!]
		\begin{center}
			\resizebox{1.\textwidth}{!}{\begin{tabular}{l|ccccccccccc}
					& $d_5$ & $d_5^2$ & $d_6$ & $d_5^3$ & $d_5\times d_6$ & $d_7$ & $d_5^4$ & $d_5^2\times d_6$ & $d_6^2$ & $d_5\times d_7$ & $d_8$\\
					\toprule
					$d_{\leq 4}$ (bosonic) &  &  & $\gtick$~\cite{Jenkins:2013zja} &  &   &  &  &  & $\gtick$~\cite{Chala:2021pll} &  & $\gtick$~\cite{DasBakshi:2022mwk} \\
					$d_{\leq 4}$ (fermionic) &  &  & $\gtick$~\cite{Jenkins:2013zja} &  &   &  &  &  & $\rtick$ &  & $\rtick$ \\
					$d_5$ & $\gtick$~\cite{Chankowski:1993tx,Babu:1993qv,Antusch:2001ck} &  &  &  & $\gtick$~\cite{Chala:2021juk} & $\otick$~\cite{Chala:2021juk} &  &  &  & & \\
					$d_6$ (bosonic) &  & $\gtick$~\cite{Davidson:2018zuo} & $\gtick$~\cite{Jenkins:2013zja,Jenkins:2013wua,Alonso:2013hga} & & & &  & \colorbox{gray}{\textcolor{white}{This\, work}} & $\gtick$~\cite{Chala:2021pll} & \colorbox{gray}{\textcolor{white}{This\, work}} & $\gtick$~\cite{DasBakshi:2022mwk} \\
					$d_6$ (fermionic) &  & $\gtick$~\cite{Davidson:2018zuo} & $\gtick$~\cite{Jenkins:2013zja,Jenkins:2013wua,Alonso:2013hga,Alonso:2014zka} & & & & & $\rtick$ &  & $\rtick$ & $\rtick$\\
					$d_7$ &  &  & & $\otick$~\cite{Chala:2021juk} & $\otick$~\cite{Chala:2021juk} & $\gtick$~\cite{Liao:2016hru,Liao:2019tep}\\
					$d_8$ (bosonic) &  &  & & & & & \colorbox{gray}{\textcolor{white}{This\, work}} & \colorbox{gray}{\textcolor{white}{This\, work}} & $\gtick$~\cite{Chala:2021pll} & \colorbox{gray}{\textcolor{white}{This\, work}} & $\gtick$~\cite{DasBakshi:2022mwk}\\
					$d_8$ (fermionic) &  &  & & & & & $\rtick$ & $\rtick$ & $\rtick$ & $\rtick$ & $\otick$~\cite{AccettulliHuber:2021uoa}
					\\\bottomrule
			\end{tabular}}
			\caption{\it State of the art of the SMEFT renormalisation (adapted from Refs.\cite{DasBakshi:2022mwk,Chala:2021pll}). The rows show the renormalised operators (categorised by dimensions and statistics). The columns show the operators contributing to RG running. Blank entries vanish, $\gtick$ denotes that the complete contribution is available, $\otick$ implies that only (but substantial) partial results are present, and $\rtick$ indicates that nothing, or very little, is known. The contribution made in this paper is marked by $\colorbox{gray}{This work}$.}\label{tab:summary}
		\end{center}
	\end{table}

	Throughout the last years and with the effort of many authors, the different contributions to the running of the Wilson coefficients of the SMEFT have been studied and revealed part by part. With Eqs.~\ref{eq:rges8} -- \ref{eq:rgecphid}, we have computed in this paper for the first time the last set of operators to be included in the running of bosonic operators at order \order{4}. Now that the LNVs\ contribution to the RGEs is computed, all the possible contributions to bosonic operators of the SMEFT are available, as shown in Table \ref{tab:summary}. There are some general remarks about the anomalous dimension matrices of the operators considered here that also apply in this case and are worth mentioning.

	All the bosonic operators being renormalised arise at tree-level in the UV completions of SMEFT. All the bosonic operators that are loop-generated (those with less than four Higgs fields) are not renormalised for the reasons mentioned in Section \ref{sec:sortingcontribs}. We have eliminated all the vanishing divergences due to irrelevant or trivial amplitudes (caused, for example, by extra loop suppression or the absence of LNV operators with gluon field strength) and those being cancelled out by symmetries (like B-only operators). The remaining coefficients are renormalised, and their equations' contributions are sorted out. In particular, we have found that four insertions of Weinberg operators can only renormalise the $\phi^8$ operator, and they are the only contribution to 1PI diagrams for this class of operators. On the other hand, the rest of the operators show at least one term containing the insertion of one $d_7$ operator.
	
	We find it interesting that most of the anomalous dimension matrix terms for operators with more than four Higgs legs (both of $d_6$ and $d_8$) deviate from the naive dimensional analysis expectation order $\mathcal{O}(\gamma)\sim 1$. Table \ref{tab:anomalousdim} shows that the insertion of a $d_6$ operator is the contribution with the largest numerical coefficients in magnitude. More specifically, operators of class $\phi^6D^2$ are enhanced by factors beyond $\mathcal{O}(\gamma)\sim 50$ from the insertion of two Weinberg operators and $\mathcal{O}_{\phi\ell}^{(1,3)}$.
	
	We have discussed that the positivity bounds impose restrictions on $d_8$ WCs, and these restrictions could be translated to LNV operators. We have shown one such case, where the constraints on the WCs of $ \mathcal{O}_{\ell \phi }^{(5)}$ and $\mathcal{O}_{\ell\phi D }^{(2)}$ are deduced using the RGE equation of $ c_{\phi^4}^{(2)} $ and its positivity bound. Assuming they are single-flavoured, these $ d_5 $ and $ d_7 $ WCs must have identical signs in any UV completion that does not generate $ \phi^4 D^4 $ at tree-level. We have inspected the constraint in models extended with heavy fermions where these LNVs are generated at tree-level, but the $ \phi^4 D^4 $ class operators are absent. We have found that these bounds are respected in Type-I and III seesaw models. We also have shown that the T-parameter bounds impact the LNV space on account of the RGEs. To elaborate on this, we have derived the restrictions on the space of Weinberg-like operators based on the T-parameter bound. To quantify the impact, we have shown in Fig.~\ref{fig:Plot2} that these put restrictions on these Wilson coefficients complementary to those imposed by neutrino masses.

	\section*{Acknowledgments}
	We thank Mikael Chala for the suggestions and discussions. We also thank José Santiago and Pablo Olgoso for help with MatchMakerEFT \cite{Carmona:2021xtq}, and Guilherme Guedes, Anisha and Maria Ramos for comments on the manuscript. This work is partly supported by SRA (Spain) under Grant No.\ PID2019-106087GB-C21 / 10.13039/501100011033 and PID2021-128396NB-100;
	by the Junta de Andalucía (Spain) under Grants No.\ FQM- 101, A-FQM-467-UGR18, and P18-FR-4314 (FEDER).
	ADC is also supported by the Spanish MINECO under the FPI programme.
	\clearpage
	
	\appendix\label{sec:appendix}

	\section{Tables of operators}\label{app:tables}
	\begin{table}[h!]
		\begin{center}
			\resizebox{\textwidth}{!}{
				\begin{tabular}{cclcl}
					\multicolumn{5}{c}{\rclass \large{\textbf{Dimension 8}}} \\[0.2cm]
					%
					\boldmath{$\phi^8$} & \rlname $\mathcal{O}_{\phi^8}$ & \rlop $( \phi^\dagger\phi)^4$ &%
					\rlop & \rlop \\[0.2cm]
					%
					& \rhname $\mathcal{O}_{\phi^6}^{(1)}$ & \rhop $(\phi^{\dag} \phi)^2 (D_{\mu} \phi^{\dag} D^{\mu} \phi)$ &%
					\rhname $\mathcal{O}_{\phi^6}^{(2)}$ & \rhop $(\phi^{\dag} \phi) (\phi^{\dag} \sigma^I \phi) (D_{\mu} \phi^{\dag} \sigma^I D^{\mu} \phi)$ \\[0.2cm]
					\multirow{-2}{*}{\boldmath{$\phi^6 D^2$}} & \rhname \textcolor{gray}{ $\mathcal{O}_{\phi^6}^{(3)}$} & \rhop \textcolor{gray}{$(\phi^\dagger\phi)^2 (\phi^\dagger D^2\phi + \text{h.c.})$} &%
					\rhname \textcolor{gray}{ $\mathcal{O}_{\phi^6}^{(4)}$} & \rhop \textcolor{gray}{$(\phi^\dagger\phi)^2 D_\mu(\phi^\dagger\ii \overleftrightarrow{D}^\mu\phi)$} \\[0.2cm]
					%
					& \rlname $\mathcal{O}_{\phi^4}^{(1)}$ & \rlop $(D_{\mu} \phi^{\dag} D_{\nu} \phi) (D^{\nu} \phi^{\dag} D^{\mu} \phi)$ &%
					\rlname $\mathcal{O}_{\phi^4}^{(2)}$ & \rlop $(D_{\mu} \phi^{\dag} D_{\nu} \phi) (D^{\mu} \phi^{\dag} D^{\nu} \phi)$ \\[0.2cm]
					& \rlname $\mathcal{O}_{\phi^4}^{(3)}$ & \rlop $(D^{\mu} \phi^{\dag} D_{\mu} \phi) (D^{\nu} \phi^{\dag} D_{\nu} \phi)$ &%
					\rlname \textcolor{gray} {$\mathcal{O}_{\phi^4}^{(4)}$} & \rlop \textcolor{gray} {$D_\mu\phi^\dagger D^\mu\phi(\phi^\dagger D^2\phi + \text{h.c.})$} \\[0.2cm]
					& \rlname \textcolor{gray} {$\mathcal{O}_{\phi^4}^{(5)} $} & \rlop \textcolor{gray}  {$D_\mu\phi^\dagger D^\mu\phi(\phi^\dagger \ii D^2\phi + \text{h.c.})$} &%
					\rlname \textcolor{gray} {$\mathcal{O}_{\phi^4}^{(6)}$} & \rlop \textcolor{gray} {$(D_\mu\phi^\dagger \phi) (D^2\phi^\dagger D_\mu\phi) + \text{h.c.}$} \\[0.2cm] 
					& \rlname \textcolor{gray} {$\mathcal{O}_{\phi^4}^{(7)}$}  & \rlop \textcolor{gray} {$(D_\mu\phi^\dagger \phi) (D^2\phi^\dagger \ii D_\mu\phi) + \text{h.c.}$} &%
					\rlname \textcolor{gray} {$\mathcal{O}_{\phi^4}^{(8)}$} & \rlop \textcolor{gray} { $(D^2\phi^\dagger\phi) (D^2\phi^\dagger\phi)+\text{h.c.}$} \\[0.2cm]
					& \rlname \textcolor{gray} {$\mathcal{O}_{\phi^4}^{(9)}$} & \rlop \textcolor{gray} {$(D^2\phi^\dagger\phi) (\ii D^2\phi^\dagger\phi)+\text{h.c.}$} &%
					\rlname \textcolor{gray} {$\mathcal{O}_{\phi^4}^{(10)}$} & \rlop \textcolor{gray} {$(D^2\phi^\dagger D^2\phi) (\phi^\dagger\phi)$} \\[0.2cm]
					& \rlname \textcolor{gray} {$\mathcal{O}_{\phi^4}^{(11)}$} & \rlop \textcolor{gray} {$(\phi^\dagger D^2\phi) (D^2\phi^\dagger\phi)$} & %
					\rlname \textcolor{gray} {$\mathcal{O}_{\phi^4}^{(12)}$} & \rlop \textcolor{gray} {$(D_\mu\phi^\dagger \phi)(D^\mu\phi^\dagger D^2\phi) + \text{h.c.}$} \\[0.2cm]
					\multirow{-7}{*}{\boldmath{$\phi^4$}} & \rlname \textcolor{gray} {$\mathcal{O}_{\phi^4}^{(13)}$} & \rlop \textcolor{gray} {$(D_\mu\phi^\dagger \phi)(D^\mu\phi^\dagger \ii D^2\phi) + \text{h.c.}$} & %
					\rlop & \rlop \\[0.3cm]
					%
					& \rhname $O_{G^2\phi^4}^{(1)}$ & \rhop $(\phi^\dag \phi)^2 G_{\mu\nu}^A G^{A\mu\nu}$ &%
					\rhname $O_{G^2\phi^4}^{(2)}$ & \rhop $(\phi^\dag \phi)^2 \widetilde{G}_{\mu\nu}^A G^{A\mu\nu}$ \\[0.2cm]
					& \rhname $\mathcal{O}_{W^2\phi^4}^{(1)}$ & \rhop $(\phi^\dag \phi)^2 W^I_{\mu\nu} W^{I\mu\nu}$ &%
					\rhname $\mathcal{O}_{W^2\phi^4}^{(2)}$ & \rhop $(\phi^\dag \phi)^2 \widetilde W^I_{\mu\nu} W^{I\mu\nu}$ \\[0.2cm]
					& \rhname $\mathcal{O}_{W^2\phi^4}^{(3)}$ & \rhop $(\phi^\dag \sigma^I \phi) (\phi^\dag \sigma^J \phi) W^I_{\mu\nu} W^{J\mu\nu}$ &%
					\rhname $\mathcal{O}_{W^2\phi^4}^{(4)}$ & \rhop $(\phi^\dag \sigma^I \phi) (\phi^\dag \sigma^J \phi) \widetilde W^I_{\mu\nu} W^{J\mu\nu}$ \\[0.2cm]
					& \rhname $\mathcal{O}_{WB\phi^4}^{(1)}$ & \rhop $ (\phi^\dag \phi) (\phi^\dag \sigma^I \phi) W^I_{\mu\nu} B^{\mu\nu}$ &%
					\rhname $\mathcal{O}_{WB\phi^4}^{(2)}$ & \rhop $(\phi^\dag \phi) (\phi^\dag \sigma^I \phi) \widetilde W^I_{\mu\nu} B^{\mu\nu}$ \\[0.2cm]
					\multirow{-5}{*}{\boldmath{$X^2 \phi^4$}} & \rhname $\mathcal{O}_{B^2\phi^4}^{(1)}$ & \rhop $ (\phi^\dag \phi)^2 B_{\mu\nu} B^{\mu\nu}$ &%
					\rhname $\mathcal{O}_{B^2\phi^4}^{(2)}$ & \rhop $(\phi^\dag \phi)^2 \widetilde B_{\mu\nu} B^{\mu\nu}$ \\[0.2cm]
					%
					& \rlname $\mathcal{O}_{W\phi^4D^2}^{(1)}$ & \rlop $\text{i}(\phi^{\dag} \phi) (D^{\mu} \phi^{\dag} \sigma^I D^{\nu} \phi) W_{\mu\nu}^I$ &%
					\rlname $\mathcal{O}_{W\phi^4D^2}^{(2)}$ & \rlop $\text{i}(\phi^{\dag} \phi) (D^{\mu} \phi^{\dag} \sigma^I D^{\nu} \phi) \widetilde{W}_{\mu\nu}^I$ \\[0.2cm]
					& \rlname $\mathcal{O}_{W\phi^4D^2}^{(3)}$ & \rlop $\text{i}\epsilon^{IJK} (\phi^{\dag} \sigma^I \phi) (D^{\mu} \phi^{\dag} \sigma^J D^{\nu} \phi) W_{\mu\nu}^K$ &%
					\rlname $\mathcal{O}_{W\phi^4D^2}^{(4)}$ & \rlop $\text{i}\epsilon^{IJK} (\phi^{\dag} \sigma^I \phi) (D^{\mu} \phi^{\dag} \sigma^J D^{\nu} \phi) \widetilde{W}_{\mu\nu}^K$ \\[0.2cm]
					& \rlname \textcolor{gray}{$\mathcal{O}_{W\phi^4 D^2}^{(5)}$} & \rlop \textcolor{gray}{$(\phi^\dag \phi) D_\nu W^{I \mu\nu}(D_\mu\phi^\dagger \sigma^I \phi + \text{h.c.}) $ } & %
					\rlname \textcolor{gray}{$\mathcal{O}_{W\phi^4 D^2}^{(6)}$} & \rlop \textcolor{gray}{$(\phi^\dag \phi) D_\nu W^{I \mu\nu}(D_\mu\phi^\dagger \text{i} \sigma^I \phi + \text{h.c.})$} \\[0.2cm]
					& \rlname \textcolor{gray}{$\mathcal{O}_{W\phi^4 D^2}^{(7)}$} & \rlop \textcolor{gray}{$\epsilon^{IJK} (D_\mu \phi^\dag \sigma^I \phi) (\phi^\dag \sigma^J D_\nu \phi) W^{K \mu\nu} $} &%
					\rlop & \rlop \\[0.2cm]
					& \rlname $\mathcal{O}_{B\phi^4D^2}^{(1)}$ & \rlop $\text{i}(\phi^{\dag} \phi) (D^{\mu} \phi^{\dag} D^{\nu} \phi) B_{\mu\nu}$ &%
					\rlname $\mathcal{O}_{B\phi^4D^2}^{(2)}$ & \rlop $\text{i}(\phi^{\dag} \phi) (D^{\mu} \phi^{\dag} D^{\nu} \phi) \widetilde{B}_{\mu\nu}$ \\[0.2cm]
					\multirow{-6}{*}{\boldmath{$X\phi^4D^2$}} & \rlname \textcolor{gray}{$\mathcal{O}_{B\phi^4D^2}^{(3)}$} & \rlop \textcolor{gray}{$ (\phi^{\dag} \phi) D_{\nu} B^{\mu\nu} (D_\mu \phi^\dagger \text{i} \phi + \text{h.c.})$} &%
					\rlop & \rlop
			\end{tabular}}
		\end{center}
		\caption{$d_8$ bosonic operators with four or more Higgs fields in the Green's basis of Ref.~\cite{Chala:2021cgt}. Operators in gray are redundant. \label{tab:dim8ops1}}
	\end{table}
	
	\clearpage
	
	\begin{table}[h!]
		\begin{center}
			\resizebox{\textwidth}{!}{
				\begin{tabular}{cclcl}
					\multicolumn{5}{c}{\bclass \large{\textbf{Dimension 7}}} \\[0.2cm]
					%
					\boldmath{$\psi^2 \phi^4$} & \blname $ \mathcal{O}_{\ell \phi}^{(7)}$  & \blop $\epsilon_{ij}\epsilon_{mn}(\ell^iC\ell^m)(\phi^j \phi^n)(\phi^\dagger \phi)$ & %
					\blop & \blop \\[0.2cm]
					%
					\boldmath{$\psi^2 \phi^2 D^2$} & \bhname $  \mathcal{O}_{\ell \phi D}^{(1)} $  & \bhop$\epsilon_{ij}\epsilon_{mn}\ell^iC(D^\mu \ell^j)\phi^m (D_\mu\phi^n)$ & %
					\bhname $  \mathcal{O}_{\ell\phi D}^{(2)} $ & \bhop $\epsilon_{im}\epsilon_{jn}\ell^iC(D^\mu \ell^j)\phi^m (D_\mu\phi^n)$ \\[0.2cm]
					%
					\boldmath{$\psi^2 \phi^3 D$} &  \blname$ \mathcal{O}_{\ell\phi De}$ & \blop $\epsilon_{ij}\epsilon_{mn}(\ell^iC\gamma_\mu e) \phi^j\phi^m D^\mu \phi^n$ & %
					\blop & \blop \\[0.2cm]
					%
					\boldmath{$\psi^2 \phi^2 X$} & \bhname $\mathcal{O}_{\ell\phi B} $ & \bhop $\epsilon_{ij}\epsilon_{mn}\ell^iC(\sigma_{\mu\nu}\ell^m)\phi^j\phi^n B^{\mu\nu}$ & 
					\bhname $ \mathcal{O}_{\ell\phi W} $ & \bhop $\epsilon_{ij}(\tau^I\epsilon)_{mn}\ell^iC(\sigma_{\mu\nu}\ell^m)\phi^j\phi^n W^{I\mu\nu}$ \\[0.2cm]
					%
					\multicolumn{5}{c}{\gclass \large{\textbf{Dimension 6}}} \\[0.2cm]
					\boldmath{$\phi^6$} & \glname $\mathcal{O}_{\phi}$ & \glop $ (\phi^\dagger \phi)^3 $ &%
					\glop & \glop \\[0.2cm]
					%
					& \ghname $\mathcal{O}_{\phi \square} $ & \ghop $ (\phi^\dagger \phi)\Square (\phi^\dagger \phi) $ &%
					\ghname $\mathcal{O}_{\phi D} $ & \ghop $ (\phi^\dagger D^\mu \phi)^\dagger (\phi^\dagger D_\mu \phi) $ \\[0.2cm]
					\multirow{-2}{*}{\boldmath{$\phi^4 D^2$}} & \ghname \textcolor{gray}{$\mathcal{O}_{\phi D}^\prime $} & \ghop \textcolor{gray} {$ (\phi^\dagger \phi) (D_\mu \phi)^\dagger (D^\mu \phi) $} &
					\ghname \textcolor{gray}{$\mathcal{O}_{\phi D}^{\prime \prime} $} & \ghop \textcolor{gray}{$ \text{i} (\phi^\dagger \phi) D_\mu (\phi^\dagger  D^\mu \phi - D^\mu \phi^\dagger \phi) $} \\[0.2cm]
					%
					& \glname $\mathcal{O}_{u\phi}$ & \glop $(\phi^\dagger \phi) \overline{q} \widetilde{\phi} u $ & %
					\glname $\mathcal{O}_{d\phi}$ & \glop $(\phi^\dagger \phi) \overline{q} \phi d $ \\[0.2cm]
					\multirow{-2}{*}{\boldmath{$\psi^2 \phi^3 $}} & \glname $\mathcal{O}_{e\phi}$ & \glop $(\phi^\dagger \phi) \overline{\ell} \phi e $ &%
					\glop & \glop \\[0.2cm]
					%
					& \ghname $\mathcal{O}^{(1)}_{Hq}$ & \ghop $ \ii(\overline{q} \gamma^\mu q) (\phi^\dagger D_\mu \phi - D_\mu \phi^\dagger \phi) $ &%
					\ghname $\mathcal{O}^{(3)}_{Hq}$ & \ghop $ \ii(\overline{q} \sigma^I \gamma^\mu q) (\phi^\dagger \sigma^I D_\mu \phi - D_\mu \phi^\dagger \sigma^I \phi) $ \\[0.2cm]
					& \ghname $\mathcal{O}_{Hu}$ & \ghop $ \ii(\overline{u} \gamma^\mu u) (\phi^\dagger D_\mu \phi - D_\mu \phi^\dagger \phi) $ &%
					\ghname $\mathcal{O}_{Hd}$ & \ghop $ \ii(\overline{d} \gamma^\mu d) (\phi^\dagger D_\mu \phi - D_\mu \phi^\dagger \phi) $ \\[0.2cm]
					& \ghname $\mathcal{O}^{(1)}_{H\ell}$ & \ghop $ \ii(\overline{\ell} \gamma^\mu \ell) (\phi^\dagger D_\mu \phi - D_\mu \phi^\dagger \phi) $ &%
					\ghname $\mathcal{O}^{(3)}_{H\ell}$ & \ghop $ \ii(\overline{\ell} \sigma^I \gamma^\mu \ell) (\phi^\dagger \sigma^I D_\mu \phi - D_\mu \phi^\dagger \sigma^I \phi) $ \\[0.2cm]
					\multirow{-4}{*}{\boldmath{$\psi^2 \phi^2 D$}}  & \ghname $\mathcal{O}_{He}$ & \ghop $ \ii(\overline{e} \gamma^\mu e) (\phi^\dagger D_\mu \phi - D_\mu \phi^\dagger \phi) $ &%
					\ghname $\mathcal{O}_{Hud}$ & \ghop $ \ii(\overline{u} \gamma^\mu d) (\phi^\dagger D_\mu \phi - D_\mu \phi^\dagger \phi) $ \\[0.2cm] 
					%
					\multicolumn{5}{c}{\oclass \large{\textbf{Dimension 5}}} \\[0.2cm]
					\boldmath{$\psi^2 \phi^2$} & \olname $ \mathcal{O}_{\ell \phi}^{(5)}$  & \multicolumn{3}{l}{\olop $\epsilon_{ij}\epsilon_{mn}(\ell^iC\ell^m)(\phi^j \phi^n)$} \\[0.2cm]
			\end{tabular}}
		\end{center}
		\caption{$d_5$, $d_6$ and $d_7$ from bases in Refs.~\cite{PhysRevLett.43.1566,Gherardi:2020det,Lehman:2014jma}, needed for the renormalisation of $d_8$ bosonic operators.}\label{tab:op67}
	\end{table}

	\bibliographystyle{style}
	\bibliography{refs} 
	
\end{document}